\documentclass[twocolumn,epsf,psfig]{revtex4-1}
\usepackage{amssymb}
\usepackage{epsfig}
\DeclareMathAlphabet{\pazocal}{OMS}{zplm}{m}{n}            
\usepackage{graphicx}
\usepackage{color}
\usepackage{bm}
\usepackage[normalem]{ulem}     
\usepackage{soul}
\usepackage{amsmath}
\usepackage{braket} 
\usepackage{rotating}
\usepackage{multirow}

\begin{document}
\title{Intrinsic orbital and spin Hall effects  in     
monolayer transition metal dichalcogenides }         

\author{Sayantika Bhowal} 
\email{bhowals@missouri.edu}
\affiliation{Department of Physics \& Astronomy, University of Missouri, Columbia, MO 65211, USA}    
\author{S. Satpathy}
\affiliation{Department of Physics \& Astronomy, University of Missouri, Columbia, MO 65211, USA}

\date{\today}

\begin{abstract}

 Orbital Hall effect (OHE) is the phenomenon of transverse flow of orbital moment in presence of an applied electric field. 
Solids with broken inversion symmetry are expected to exhibit a strong  OHE due to the presence of  an intrinsic orbital moment at individual momentum points in the Brillouin zone, which in presence of an applied electric field, flows in different directions causing  a net orbital Hall current. 
 Here 
 we provide a comprehensive understanding of the effect and its tunability in the monolayer 2D transition metal dichalcogenides (TMDCs). Both metallic and insulating TMDCs are investigated from full density-functional calculations, effective $d$-band tight-binding models,  as well as a minimal four-band model 
 for the  valley points that captures the key physics of the system.
For the tuning of the OHE, we examine the role of hole doping as well as the change in the band parameters,
which, e. g., can be controlled by strain. 
 We demonstrate that the OHE is a more fundamental effect than the spin Hall effect (SHE), with the momentum-space orbital moments inducing a spin moment in the presence of the spin-orbit coupling, leading to the SHE.
  The physics of the OHE, described here, is relevant for  2D materials with broken inversion symmetry in general, even beyond  the TMDCs, providing a broad platform for future research.

\end{abstract}

\maketitle

\section{Introduction}

Among the various types of Hall effects, anomalous and spin Hall effects (AHE, SHE) have received considerable attention in recent times  not only due to the fundamental physics involved, but also because of their immense applications in spintronics devices \cite{You,Bhowalnpj,Yang,BTIO,Naka}. 
An orbital counterpart of the SHE is the orbital Hall effect (OHE), where an external electric field creates an imbalance in orbital moment $\vec L$, instead of a spin imbalance in the SHE, leading to a transverse flow of orbital current.
Interestingly, the OHE is proposed to be the origin of large AHE and SHE in several materials and usually has much larger magnitude than the SHE \cite{Kotani2009,Tanaka,Kotani2008}. 
Thus, understanding of the OHE not only can guide the ongoing search for new materials with large AHE or SHE, but also can have applications in potential ``orbitronics" devices, beyond spintronics, for carrying information.   

In spite of a number of theoretical predictions, OHE has not been directly observed in experiments, partly because of the common belief of orbital quenching in solids \cite{Kittel, Go}. 
Recently Go {\it et.al.} put forward the mechanism of the OHE in centrosymmetric 3D systems, where it was shown that even though the orbital moment is quenched prior to the application of the electric field, the external electric field can induce a nonzero $\vec L$ in the momentum space, generating a large OHE \cite{Go}. The generated orbital texture by the applied electric field as well as the resulting OHE are found to be ubiquitous in multi-orbital systems \cite{Jo}.

Recently, it has been proposed that 
{\it non}-centrosymmetric systems with time-reversal ($\mathcal{T}$) symmetry, such as the transition metal dichalcogenides (TMDCs), can also host a large
OHE \cite{ohe_our,Canonico}, the mechanism of which is, however, significantly different from the centrosymmetric case \cite{Go,Jo}. 
Unlike centrosymmetric materials an intrinsic $k$-space orbital moment exists in systems with broken inversion ($\mathcal{I}$) symmetry, even prior to the application of an external electric field. These orbital moments in the $k$-space, flow in the transverse direction in response to the applied electric field $\vec E$ in the Hall effect due to the anomalous velocity $\vec v \propto \vec E \times\vec \Omega$, arising from the nonzero Berry curvature $\vec \Omega$ in the system. This flow of orbital moment, then, leads to a large OHE. The large orbital to spin Hall conductivity ratio, makes these materials particularly suitable for the detection of OHE. 

While the general principles of the OHE and the SHE in non-centrosymmetric TMDCs  were discussed in our work \cite{ohe_our} based on an effective four-band tight-binding (TB) model, valid near the valley points $K$ and $K^\prime$, 
here we present an in-depth study of these effects by considering a general nearest neighbor (NN) TB model for the transition metal $d$-orbitals, describing the entire Brillouin zone (BZ). Using this full TB model, we discuss the 
crucial roles of the inter-orbital hopping, electron occupation of $d$-orbitals and the strength of the spin-orbit coupling (SOC) in both OHE and SHE, providing the possible knobs to control these effects in monolayer TMDCs.
 The results of the model calculations are further corroborated by the density functional calculations, employed to study insulating (MX$_2$; M = Mo, W and X = S, Se, Te) as well as metallic (NbS$_2$) TMDCs. The  mechanism is also applicable to other 2D systems, where the inversion symmetry is absent.
%
%



Before going into details, we discuss here the basic physical picture of the OHE in TMDCs.
TMDCs lack the $\mathcal{I}$ symmetry which plays the key role in generating the intrinsic orbital moment in the k-space. The broken $\mathcal{I}$ symmetry induces new inter-orbital hoppings in the system that hybridize between different $d$-orbitals. As a result, although the individual cubic harmonics, e.g., $|d_{x^2-y^2} \rangle$ and $|d_{xy} \rangle$, do not carry any orbital angular momentum, the hybridized orbitals, such as ($|d_{x^2-y^2} \rangle \pm i |d_{xy}\rangle)$ can have a finite orbital moment. In TMDCs, the orbital moments are generated around the two valley points ($K$, $K^\prime$) which have opposite directions due to the presence of $\mathcal{T}$ symmetry that dictates $\vec M (\vec k) = -\vec M (-\vec k)$. 
In order to generate the OHE, 
these orbital moments need to flow in opposite directions in real space in response to an applied electric field in the Hall effect. This indeed is the case for TMDCs as the two valleys carry Berry curvatures with opposite signs. As a result, the electrons at the two valleys acquire anomalous velocity with opposite directions, leading to a large intrinsic OHE.
Clearly the OHE can occur without SOC. However,
 the presence of a sizable SOC in TMDCs couple the orbital moments to the spin moments, which gives rise to the well known  ``valley dependent spin splitting". Furthermore, these spin polarized bands have nonzero spin Berry curvatures, leading to the SHE, which is however significantly smaller in magnitude as compared to the OHE. 


 The remaining parts of the paper are organized as follows. The detail of the crystal structure of TMDCs and the computational methods, used in the present work are described in Section II. This is followed by Section III, where we discuss a general NN TB model for the $d$-orbitals of the transition metal atom M, relevant for the 2H-MX$_2$ structure. Here, we first qualitatively analyze the role of the broken $\mathcal{I}$ symmetry in inducing additional inter-orbital hoppings. With these effective $d$-$d$ hoppings into account, we illustrate the generation of the $k$-space orbital moment and the OHE, and how they can be manipulated by hole doping or 
 by engineering the band structure. 
 This is followed by our study of the effect of the spin-orbit coupling on both the OHE and the SHE in Section IV. We then construct
an effective four-band model for the valley points from this full $d$ orbital NN TB model in Section V, which provides useful insight into the tuning of both OHE and SHE by hole doping.  
  The results of the TB model are also corroborated by  explicit calculations based on density functional theory (DFT) for a series of metallic and insulating TMDCs in Section VI, and the results are summarized in Section VII.

\section{Crystal structure and computational methods}

\begin{figure}[h] 
\centering
\includegraphics[width=\columnwidth]{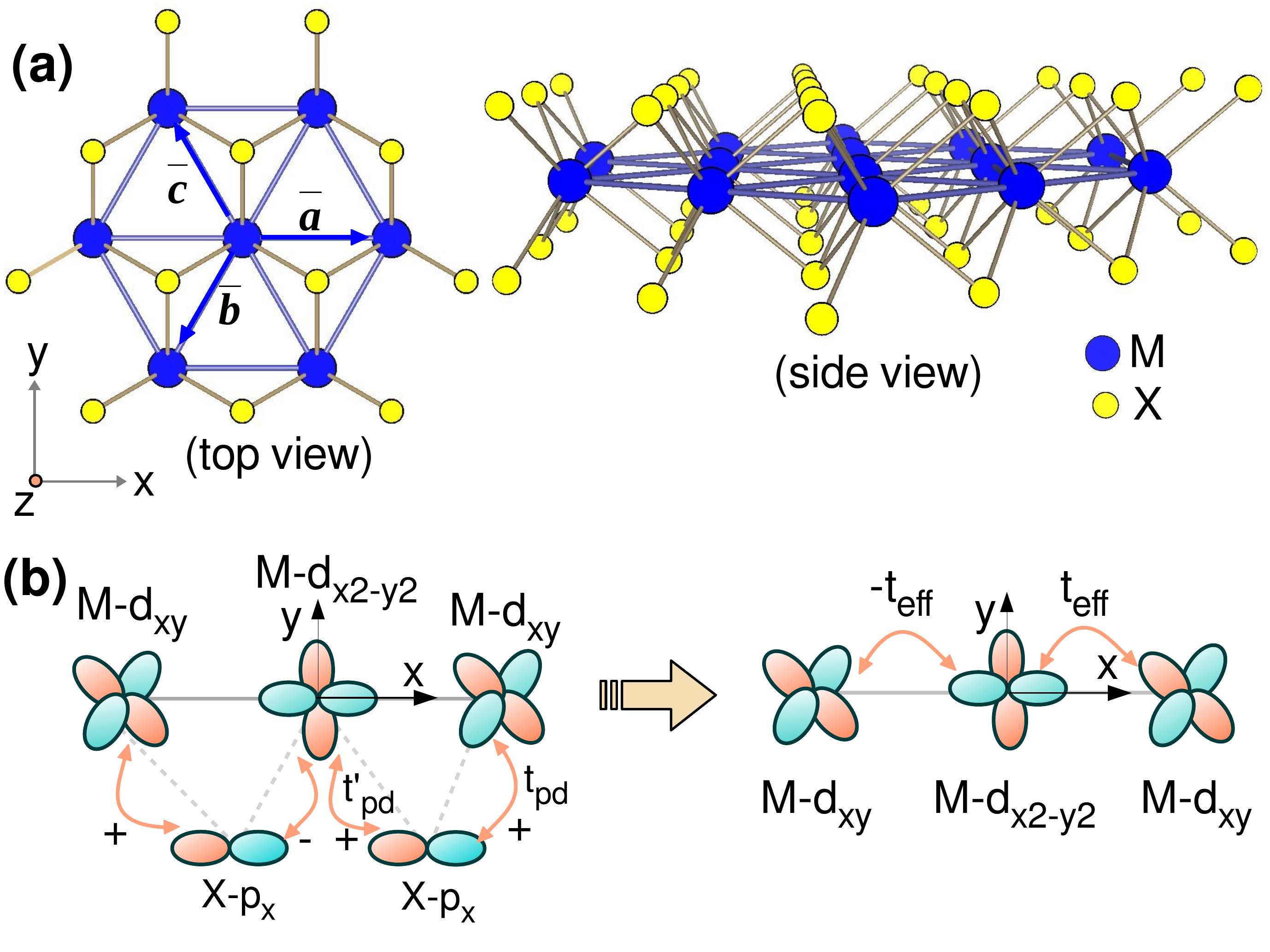}
\caption {Crystal structure and effective inter-orbital hopping. (a) The triangular network formed by the transition metal M atom as viewed from the top and from the side.
The hopping between the metal atoms along the nearest-neighbor vectors $\vec \alpha = \vec a, \ \vec b,$ and $ \vec c$, are related via the $C_3$ rotational symmetry.
(b) Direction dependence of the sign of the inter-orbital hopping between M-$d_{xy}$ and M-$d_{x^2-y^2}$ orbitals, mediated via the  X-$p$ orbitals, that break the $\cal I$ symmetry. The $\pm$ signs in (b) represent the relative signs of the $p$-$d$ hopping with the magnitude denoted by $t_{pd}$ and $t_{pd}^\prime$. 
The metal-ligand interaction leads to the sign change of the effective $d$-$d$ hopping $t_{\rm eff}$ as indicated in the right part of (b).
}
\label{fig1} 
\end{figure}

We have chosen the monolayer TMDCs 2H-MX$_2$\cite{materialscloud} for a detail study because 
it is a well-known material with  broken inversion symmetry. The crystal structure has  
the $D_{3h}$ point group symmetry, with the character table and the irreducible representations spanned by the $d$ orbitals shown in Table \ref{characterTable}. 
%
The crystal structure is indicated in Fig. \ref{fig1} (a). The unit cell contains one formula unit with a single M atom and two X atoms.The M atoms are arranged on a triangular network in the $a-b$ plane, while the two X atoms are situated out of this plane in such a way that the structure has the horizontal mirror symmetry $\sigma_h$ but not the $\mathcal{I}$ symmetry. 
The structure also has the $C_3$ rotational symmetry.
As seen from Fig. \ref{fig1} (a), each M atom is surrounded by six X atoms, forming an MX$_6$ trigonal prism. 
Due to the triangular network formed by the M atoms, each M atom has six M atoms as its nearest neighbors along $\pm \vec a, \pm \vec b$, and $\pm \vec c$ directions [see Fig. \ref{fig1}, (a)]. Here, $ \vec a = a \hat x,
~ \vec b = - (a/2)  ( \hat x + \sqrt{3} \hat y)$ and 
$ \vec c = - (a/2)  ( \hat x - \sqrt{3} \hat y)$, $a$ being the lattice constant and  the distance between the NN metal atoms.

Density-functional calculations presented here were calculated using the Quantum Espresso \cite{QE} 
as well as the muffin-tin orbitals based (NMTO) \cite{nmto} codes.
 %
 %
 All  calculations were performed with the relaxed structure, obtained by relaxing the atomic positions until the Hellman-Feynman forces on each atom becomes less than 0.01 eV/\AA. 
%
The k-space orbital moment of the monolayer was calculated using Quantum Espresso and Wannier90 codes \cite{QE, w90_code}. 
The {\it ab-initio} wave functions of the ground state,  computed within a self-consistent calculation, are used to construct the maximally-localized Wannier functions \cite{MLWF} employing the Wannier90 code \cite{w90_code}.
The wannierisation process was converged to 10$^{-10}$ \AA$^2$ having an average spread less that $\sim$ 1 \AA$^2$ of the Wannier functions. The orbital moment in the k-space is, therefore, calculated in the Wannier space \cite{w90}. 

The complementary muffin-tin orbitals based method (NMTO) \cite{nmto} was also used to compute the k-space orbital moment. In addition, NMTO was used to calculate the orbital and the spin Hall conductivities (OHC/ SHC) in the following manner. First, effective TB hopping matrix elements between the M-$d$ orbitals were obtained with several neighbors,  which yielded the full TB Hamiltonian valid everywhere in the BZ. 
Using this full TB model, we obtained the eigenvalues and wave functions, from which 
the orbital and the spin Hall conductivities were computed from a Brillouin zone sum,
 by summing over  $400 \times 400$ $k$ mesh points in the 2D zone.
The computed orbital moments using the Wannier90 or the NMTO method agree well with each other.

\section{Tight-binding results: Orbital moment and orbital Hall Effect } \label{TB}

In this Section, we construct a nearest-neighbor (NN) TB model Hamiltonian for the metal $d$ orbitals  in the monolayer 2H-MX$_2$ structure, and using this model Hamiltonian, we describe the $k$-space orbital moment and the resulting orbital Hall effect and
 how they are affected by the parameters of the Hamiltonian as well as by doping. 

\subsection{Broken $\mathcal{I}$ symmetry induced inter-orbital hoppings }
The broken inversion  symmetry $\mathcal{I}$ plays an important role in the electronic structure  as well as 
in the OHE of the TMDCs. It leads to a mixing of the orbitals with complex coefficients in the wave functions
producing thereby orbital moments at the various points in the Brillouin zone, which then lead to a robust OHE.
The chalcogen X atoms break the symmetry and introduce new inter-orbital hoppings between the neighboring M atoms, which were zero without the broken symmetry.  
The chalcogen orbitals must be kept in the TB model to incorporate the effect of the broken $\mathcal{I}$ symmetry. However, the effect can be included via the L\"owdin downfolding method \cite{downfolding} to produce an effective M $d-d$ hopping, eliminating the chalcogen orbitals from the basis set.
A detail description of this point is given in  Appendix A, but below we describe the essential physics from perturbation theory.

Using the perturbation theory, the effective $d-d$ hoppings and their sign change with the direction of hopping
may be inferred. These  hoppings are mediated via the X-$p$ orbitals with a magnitude 
$t_{\rm eff} \sim  t_{pd}t_{pd}^\prime/\Delta_{pd}$, where  $t_{pd}$ and $t_{pd}^\prime$ are the hoppings between the metal atoms and the intermediate ligand atom as indicated in Fig. \ref{fig1} (b), and 
$\Delta_{pd}$ is the energy difference between the metal and ligand orbitals. 
Considering first the ligand $p_x$ orbitals, the metal-ligand hopping $t_{pd}$ is given by the Slater-Koster Tables \cite{SlaterKoster}: 
$t_{x,x^2-y^2} = 2^{-1}3^{1/2}l(l^2 - m^2)V_{pd \sigma}+l(1-l^2 + m^2)V_{pd \pi}$ and
$t_{x,xy} = 3^{1/2}l^2m V_{pd \sigma}+ m(1-2l^2)V_{pd \pi}$, where ($l, m, n$) are the direction
cosines of the ligand with respect to the metal atom, and $V_{pd \sigma}$ and $V_{pd \pi}$
are the usual Slater-Koster hopping parameters. For the orbitals shown in \ref{fig1} (b), these expressions lead
to the $p-d$ hopping which are of the same magnitude, but different signs as indicated. The same goes for the 
$p-d$ hopping for the $p_y$ orbitals. Thus, the perturbation expression stated above leads to a change of sign 
of $t_{\rm eff}$ for hopping to left or right as indicated in \ref{fig1} (b).
As elaborated in Appendix B, such inter-orbital hopping with a sign change eventually leads to the formation of complex orbitals $d_{x^2-y^2} \pm i d_{xy}$ near the valley points, which carry a non-zero orbital moment $l_z = \pm 2\hbar$.
As an aside, we note that a similar sign change with respect to the direction of hopping
leads to the  well-known Rashba effect \cite{Rashba}, which arises from the broken mirror symmetry.

\subsection{Tight-binding model  Hamiltonian}\label{alld}
We now construct a general NN TB model on the triangular lattice for the M-$d$ orbitals taking into account the effective $d$-$d$ hoppings, as discussed above. This NN model for the five TM-$d$ orbitals (labeled by $m$) on site $i$ with the 
field operators $c_{im}$ and $c^\dagger_{im}$  provide an overall reasonable description of the 
electronic structure of the TMDCs over the entire BZ. 
The TB Hamiltonian, written in the Bloch function basis
\begin{equation}\label{Bloch}
 c^\dagger_{\vec{k} m} = \frac{1}{\sqrt{N}} \sum_{i} e^{i \vec{k}\cdot \vec{R}_i} c^\dagger_{im},
\end{equation}
with $\vec{k}$ being the Bloch momentum and 
 the order of the basis set in the sequence ($d_{xy}, d_{x^2-y^2}, d_{3z^2-r^2}, d_{xz}, d_{yz}$) is given by
\begin{eqnarray}     \label{H}  
{\cal H} (\vec k ) &= 
\left[
{\begin{array}{*{20}c}
    h_{11} & h_{12} & h_{13} & 0 & 0  \\
    h^*_{12} & h_{22} & h_{23} & 0 & 0  \\
    h^*_{13} & h^*_{23} & h_{33} & 0 & 0 \\
    0 & 0 & 0 & h_{44} & h_{45} \\
    0 & 0 & 0 & h^*_{45}& h_{55} \\
\end{array} }  \right],
\end{eqnarray} 
where
\begin{widetext}
\begin{eqnarray} \nonumber
 h_{11} &=& \varepsilon_1+2t^a_1\cos k_a+2t^b_1(\cos k_b + \cos k_c)\\ \nonumber
 h_{22} &=& \varepsilon_1+2t^a_2\cos k_a+2t^b_2(\cos k_b + \cos k_c)\\\nonumber
 h_{33} &=& 2t^a_3(\cos k_a + \cos k_b + \cos k_c)\\ \nonumber
 h_{44} &=& \varepsilon_2+2t^a_4\cos k_a+2t^b_4(\cos k_b + \cos k_c)\\ \nonumber
 h_{55} &=& \varepsilon_2+2t^a_5\cos k_a+2t^b_5(\cos k_b + \cos k_c)\\ \nonumber
 h_{12} &=& 2it^a_6 \sin k_a + i(t^b_6+t^c_6) (\sin k_b+\sin k_c) 
 + (t^b_6-t^c_6) (\cos k_b - \cos k_c) \\ \nonumber
 h_{13}&=&  2it^a_7 \sin k_a + i(t^b_7 + t^c_7) (\sin k_b+\sin k_c) 
 + (t^b_7-t^c_7) (\cos k_b - \cos k_c) \\ \nonumber
 h_{23}&=&  2t^a_8 \cos k_a + (t^b_8+t^c_8) (\cos k_b + \cos k_c) 
 + i(t^b_8-t^c_8) (\sin k_b-\sin k_c) \\ 
 h_{45}&=&  2it^a_9 \sin k_a + i(t^b_9 + t^c_9) (\sin k_b+\sin k_c) 
 + (t^b_9-t^c_9) (\cos k_b - \cos k_c).
\end{eqnarray}
\end{widetext}

 \begin{table*} [t]    
\caption{Typical TB parameters (in meV)  for  the monolayer 2H-MX$_2$ structure. 
These were obtained using the NMTO method.
}
\centering
\setlength{\tabcolsep}{8pt}
 \begin{tabular}{ c c c c c c c c c c c}
\hline
    $\varepsilon_1$ & $\varepsilon_2$ & $t^a_1$ & $t^a_2$ & $t^a_3$ & $t^a_4$ & $t^a_5$ &  $t^a_6$ &  $t^a_7$ &  $t^a_8$ &  $t^a_9$\\
\hline\hline
732 & 2.6 $\times 10^3$ &185 & -58 & -207 & -247 & -86 & -310 & 358 & 348 & -76\\
\hline
\end{tabular} 
\label{tab1} 
\end{table*}

 \begin{table} [h]    
\caption{TB parameters (in meV) for hopping along the directions ${\vec b}$ and ${\vec c}$ derived from the parameters in Table \ref{tab1} and using the $C_3$ rotational symmetry as discussed in the text. }
\centering
\setlength{\tabcolsep}{4pt}
 \begin{tabular}{ c c c c c c c c c}
%
\hline
     $t^b_1$ & $t^b_2$ & $t^b_3$ & $t^b_4$ & $t^b_5$ &  $t^b_6$ &  $t^b_7$ &  $t^b_8$ &  $t^b_9$\\
\hline\hline
3 & 124 & -207 & -126 & -207 & -204 & 122 & -484 & -147 \\ 
\hline
    $t^c_1$ & $t^c_2$ & $t^c_3$ & $t^c_4$ & $t^c_5$ &  $t^c_6$ &  $t^c_7$ &  $t^c_8$ &  $t^c_9$\\
\hline\hline
3 & 124 & -207 & -126 & -207 & -416 & -480 & 136 & -5\\
\hline
\end{tabular} 
\label{tab3} 
\end{table}

 \begin{figure}[ht]
\centering
\centering \includegraphics[width=6.0cm]{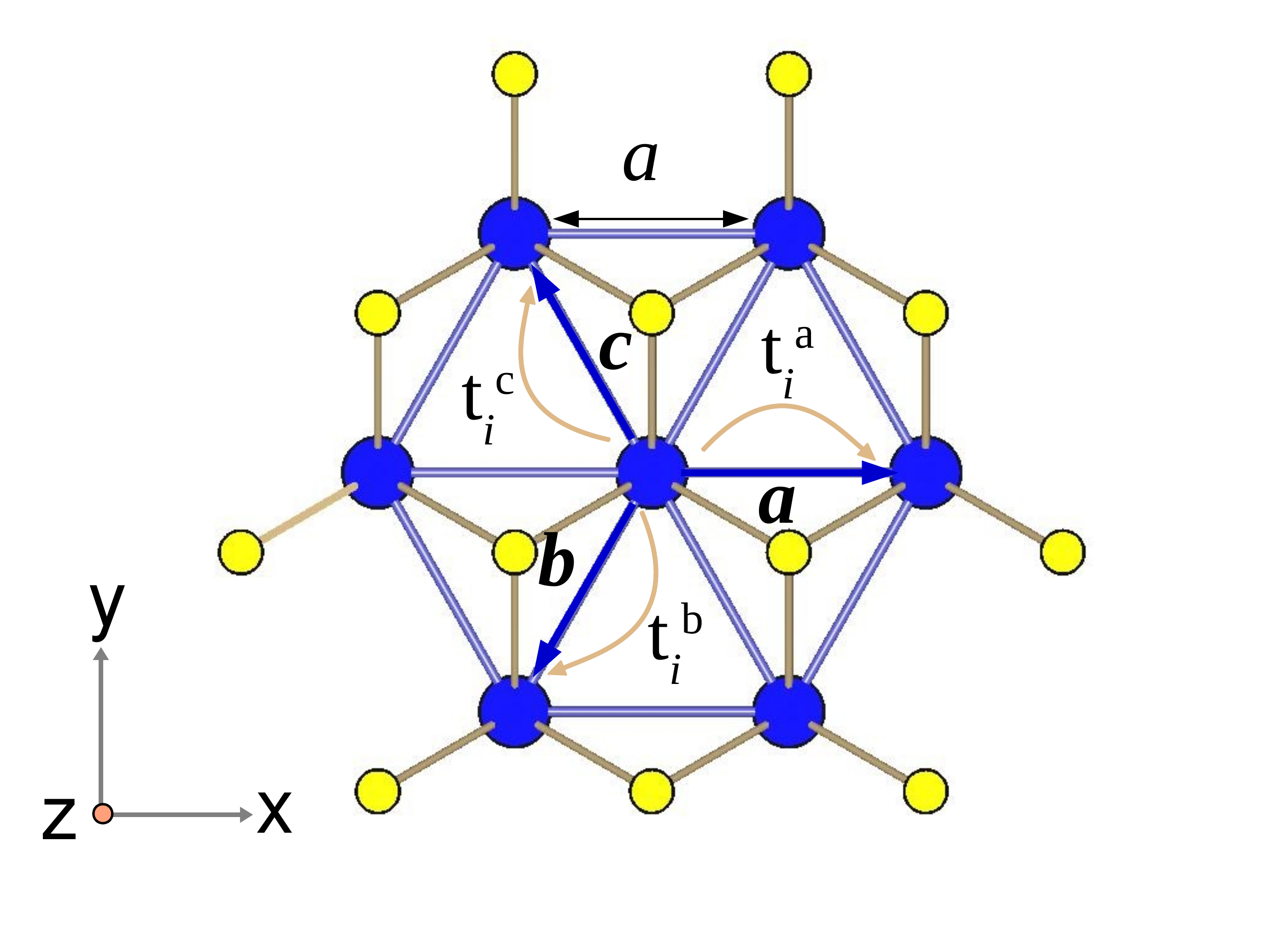}
\includegraphics[width=\columnwidth]{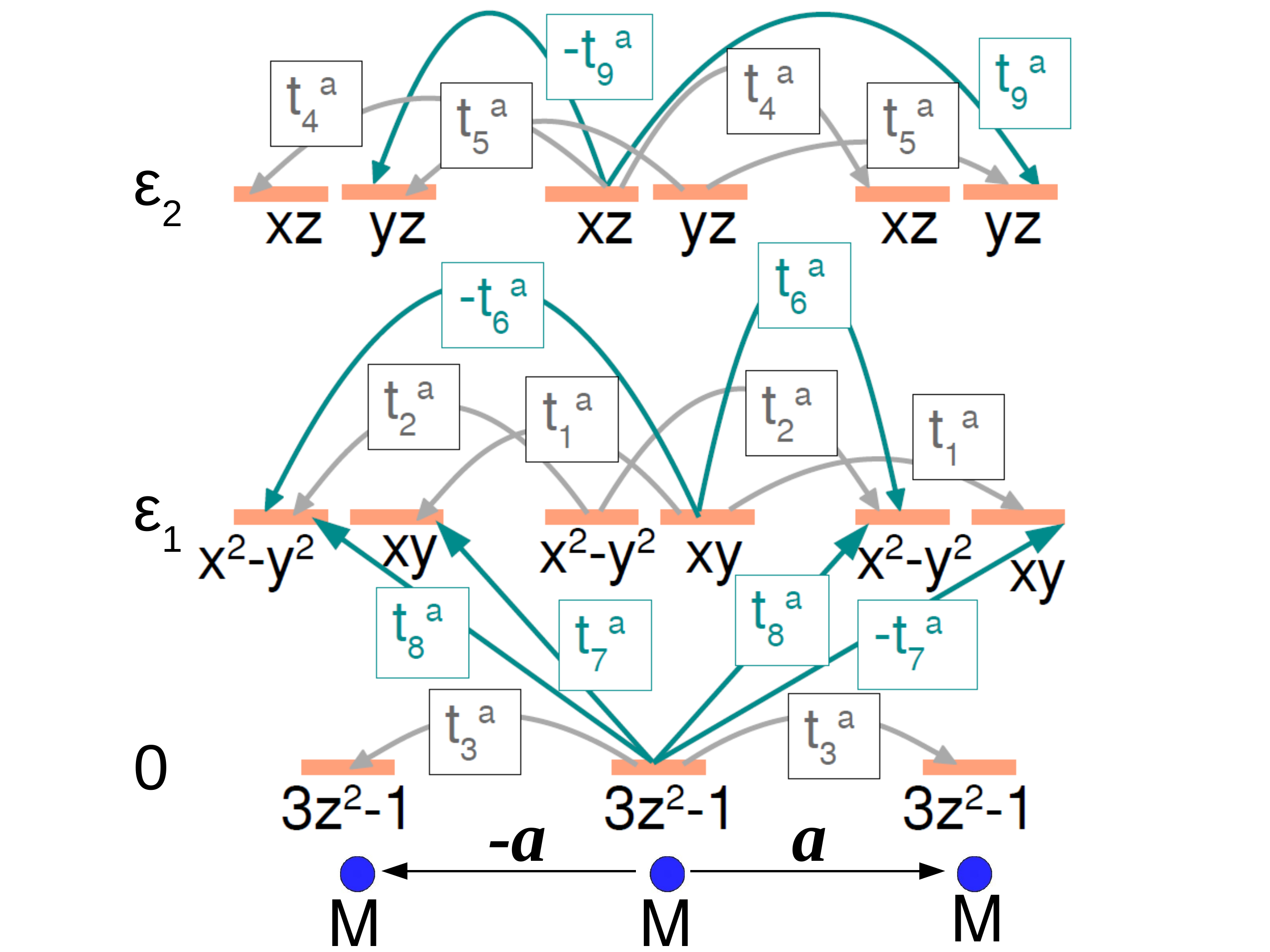}
\caption {Tight-binding parameters in the effective $d$ band model. There are eleven independent parameters:
 two onsite energies $\varepsilon_1$ and $\varepsilon_2$, and nine NN hopping parameters $t_i^a$, $i=1 - 9$ for hopping along $\vec a$. The hopping parameters along $\vec b$ and $\vec c$ can be obtained 
  from the $C_3$ symmetry, as discussed in the text. Note that some hoppings have left/right sign asymmetry 
  as illustrated in Fig. \ref{fig1}.
}
\label{fig_hop} 
\end{figure}
Here, $k_\alpha = \vec k \cdot \vec \alpha$,  $\vec \alpha = \vec a,\   \vec b,\   \vec c \ $ denote the directions  of the NN M atoms, and the TB on-site energies $\varepsilon_i$ and the NN hopping matrix elements $t_i^\alpha$ are as indicated in 
Fig. \ref{fig_hop}. 

The TB parameters for hopping along the three NNs are not independent, 
but rather they are related to each other via the $C_3$ rotational symmetry of the structure. 
Thus by knowing the hopping  $t_i^a$ for the NN M atoms along $\vec a$, we can compute the corresponding hoppings $t_i^b, t_i^c$ along $\vec b$ and $ \vec c$  respectively using the transformation relation 
\begin{equation}\label{transf}
 H^{\gamma}_{\rm hop} = R^T (\theta_\gamma) H^a_{\rm hop} R (\theta_\gamma),
\end{equation}

where the hopping matrix along $\vec a$,  $H^a_{\rm hop}$,  with the basis
 $\phi_\alpha \equiv \{ xy, x^2-y^2, 3z^2-r^2, xz, yz \}$ 
 is given by
\begin{eqnarray}     \label{Ha}  
H^{a}_{\rm hop} &= 
\left[
{\begin{array}{*{20}c}
    t_1^a & t_6^a & t_7^a & 0 & 0  \\
    -t_6^a & t_2^a & t_8^a & 0 & 0 \\
    -t_7^a & t_8^a  & t_3^a & 0 & 0 \\
    0 & 0 & 0 & t_4^a & t_9^a\\
    0 & 0 & 0 & -t_9^a & t_5^a \\
\end{array} }  \right].
\end{eqnarray} 
The rotation matrix is
\begin{eqnarray}     \label{rot}  
R (\theta_\gamma) &= 
\left[
{\begin{array}{*{20}c}
    \cos 2\theta_\gamma & \sin 2\theta_\gamma & 0  & 0 & 0  \\
    -\sin 2\theta_\gamma & \cos 2\theta_\gamma & 0 & 0 & 0 \\
    0 & 0 & 1 & 0 & 0\\
     0 & 0 & 0 & \cos \theta_\gamma & -\sin \theta_\gamma \\
   0 & 0 & 0 & \sin \theta_\gamma & \cos \theta_\gamma \\ 
\end{array} }  \right],
\end{eqnarray} 
where $\theta_\gamma = 2\pi/3$ and $4\pi/3$ correspond to the hopping along $\vec b$ and $ \vec c$ directions respectively. Using the matrices (\ref{Ha}) and (\ref{rot}) and by performing the matrix multiplication in Eq. (\ref{transf}), we can compute the hopping parameters along $\vec b$ and $ \vec c$ directions.
Thus the only unknown parameters in the TB model (\ref{H}) are $t^a_i, \epsilon_1,$ and $\epsilon_2$. 
A typical set of parameters is listed in Table \ref{tab1} for hopping along $\vec a$,  and the computed parameters for hopping along the other two NN directions are listed in
 Table \ref{tab3}.  
Note that the presence of the reflection symmetry $\sigma_h$ in the structure (see Appendix C) allows  hybridization only within the subspaces of $(xy,  x^2-y^2,  3z^2-r^2)$ and ($xz, yz$) orbitals, and these two subspaces do not intermix
as seen from the Hamiltonian Eq. (\ref{H}).


\subsection{Orbital moment  } 
\label{orbandohe}

In this Section, we discuss the presence of the intrinsic orbital moment in the TMDCs, the magnitude of which are quite large near the valley points $K, K^\prime$, which in turn play a crucial role in driving a large orbital Hall conductivity (OHC) via the orbital Berry curvature.  

Symmetry considerations for the TMDCs dictate that a non-zero orbital moment $\vec M(\vec k)$ can exist at the 
individual momentum points in the BZ, though the total moment, obtained by summing over  the BZ,
 must vanish.
For the TMDCs, we have the inversion symmetry $\mathcal{I}$ broken,  while the time reversal symmetry 
${\cal T}$  is present.
 %
If the inversion $\mathcal{I}$ and the time-reversal ${\cal T}$ symmetries are present,
the following well known conditions hold for the orbital moment for the momentum point $k$ in the 
Brillouin zone:
\begin{eqnarray} \nonumber 
 \vec M(\vec k) &=& \vec M(-\vec k) \ \ \ \ \ \ {\rm (Inversion \ {\mathcal{I}} \ present)}  \\ 
 \vec M(\vec k) &=& -\vec M(-\vec k) \ \ \ \ {\rm (Time  \ Reversal \  {\cal T} \ present)}.
\end{eqnarray}
 Therefore, in presence of both $\mathcal{I}$ and ${\cal T}$ symmetries, $\vec M(\vec k)$ vanishes at each k-point of the BZ. 
 
\begin{figure}[t]
\centering
\includegraphics[scale=.28]{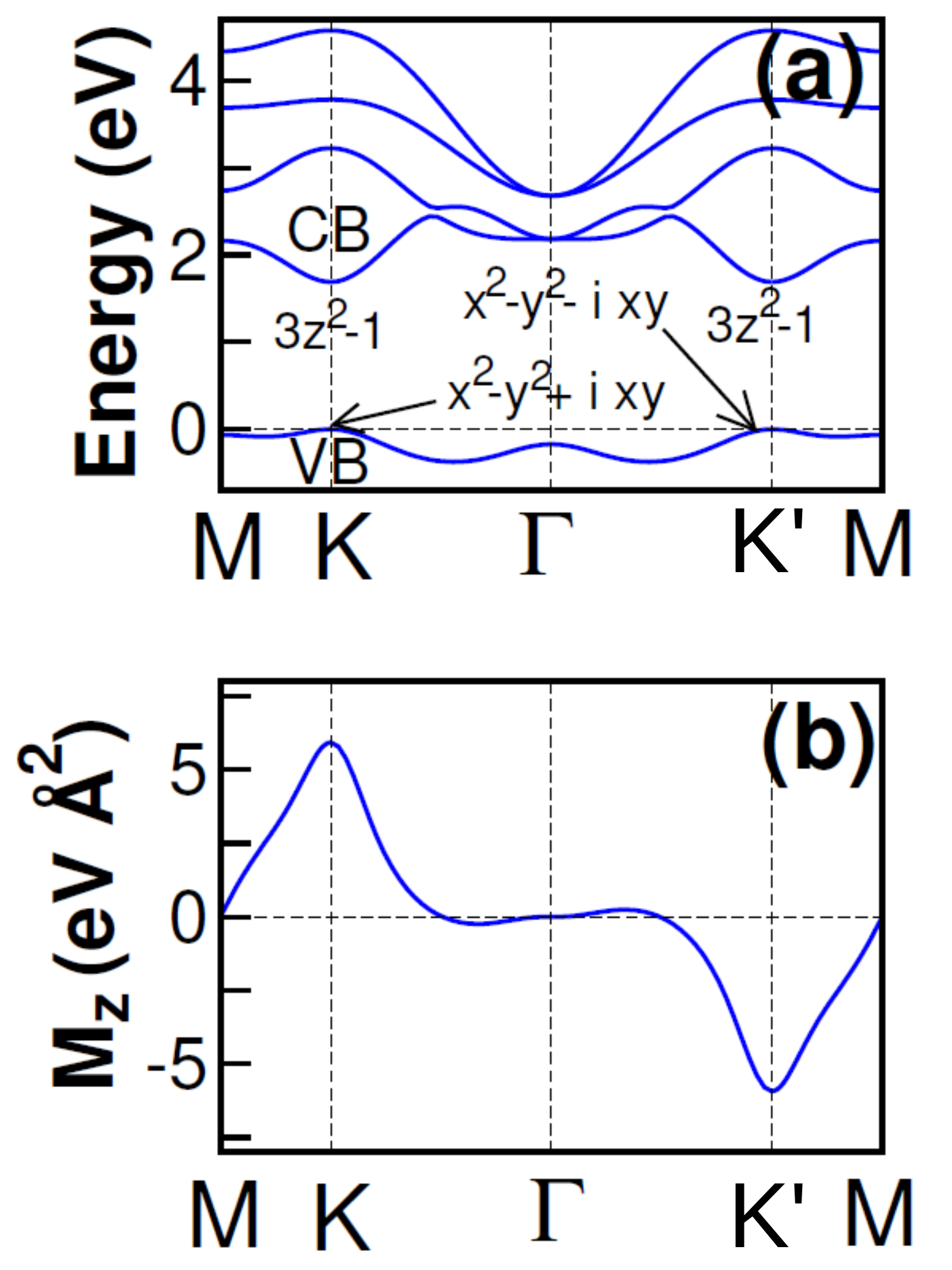}
\caption {Band structure (a) and the orbital moment $M_z (\vec k)$ for the valence band (VB) (b),
obtained for the Hamiltonian (\ref{H}) with the TB parameters of Table \ref{tab1}. 
}
\label{fig2} 
\end{figure}

 On the other hand, $\vec M(\vec k)$ is non-zero, if either of the two symmetries is broken,
 which is the case for  
the TMDCs due to the broken $\mathcal{I}$. 
 Furthermore, due to the presence of ${\cal T}$, 
 the orbital moments have opposite signs at
 the $K$  and $K'$ momentum points.
 The same symmetry condition  dictates that the orbital  moment at the $\Gamma$ point must vanish.  
The computed values $ M_z(\vec k)$,
shown in Figs. \ref{fig2} and \ref{fig3}, are consistent with these symmetry arguments,
 and they also follow the $C_3$ rotational symmetry of the structure. 

The typical band structure for the TMDCs, computed by diagonalizing the Hamiltonian (\ref{H}), is shown in Fig. \ref{fig2} (a) for the parameters listed in Table \ref{tab1}. Depending on the occupation of the M-$d$ orbitals, we have either  an insulating or a metallic state.
While for the  TM atom M (M = Mo, W) with $d^2$ configuration, the lowest band in Fig. \ref{fig2} is completely occupied resulting in an insulating state, for the TM atoms with $d^1$  configuration such as NbS$_2$, the lowest band in Fig. \ref{fig2} is only partially filled leading to a metallic state. 
In either of these cases, the valence band maximum and the conduction band minimum are both located at the two valleys $K, K^\prime$, which turn out to be the dominant points for the OHE due to the large orbital moments there.

It is easy to see that a large orbital moment is expected to occur at the valley points.
Diagonalization of the Hamiltonian (\ref{H}) at $K$ (-4$\pi$/3a, 0) and $K'$ (4$\pi$/3a, 0) points shows that the eigenvectors at those two valleys are: 
$|v \rangle = (\sqrt{2})^{-1} (|x^2-y^2 \rangle + i\tau |xy \rangle)$, $|c \rangle = |3z^2-r^2 \rangle$, and 
$|c' \rangle = (\sqrt{2})^{-1} (|x^2-y^2 \rangle - i\tau |xy \rangle)$, in ascending
order of energies, where $\tau = \pm 1$ is the valley index for $K$ and $K'$ points respectively. The first two eigenvectors corresponding to the lowest two energy eigenvalues constitute the valence and the conduction bands  at the $K$ and $K'$ points. 
The imaginary coefficients of the eigenvectors $|v \rangle$ and $|c' \rangle$, which can be traced back to the directional sign change of the effective $d-d$ hopping integral (Appendix B),  lead to a non-zero orbital moment for these states. 
As emphasized earlier, the inversion symmetry breaking was instrumental in producing the effective $d-d$ hopping in the first place, without which the orbital moment would be zero everywhere in the Brillouin zone.


The magnitude of the orbital moment $\vec M (\vec k)$  can be computed using
the modern theory of orbital moment \cite{Niu, Vanderbilt}, viz.,
\begin{eqnarray}\nonumber \label{orb}
 \vec M(\vec k) &=& -2^{-1} ~ \text {Im} [\langle \vec \nabla_k u_{\vec k} | \times ({\cal H} -\varepsilon_{\vec k}) |\vec \nabla_k u_{\vec k} \rangle ] \\ 
& & + ~\text {Im} [\langle \vec \nabla_k u_{\vec k} | \times (\epsilon_F-\varepsilon_{\vec k}) |\vec \nabla_k u_{\vec k} \rangle ],
\end{eqnarray}
where $\varepsilon_{\vec k}$ and $u_{\vec k}  $ are the energy eigenvalues and the eigen functions
for a given band. 
The total orbital moment is obtained by summing over the occupied states at a given $\vec k$ point.

The first term in Eq. (\ref{orb}) is  the angular momentum ($\vec r \times \vec v$)
 contribution due to the self-rotation of the Bloch electron wave packet, while the second term
is due to the center-of-mass  motion of the wave packet. 
Note that for 2D materials, the orbital moments  can only point in the direction normal to the plane,
which is self evident due to the presence of the cross product between two vectors that lie on the plane
of the structure in Eq. (\ref{orb}).

 \begin{figure}[t]
\centering
\includegraphics[scale=.45]{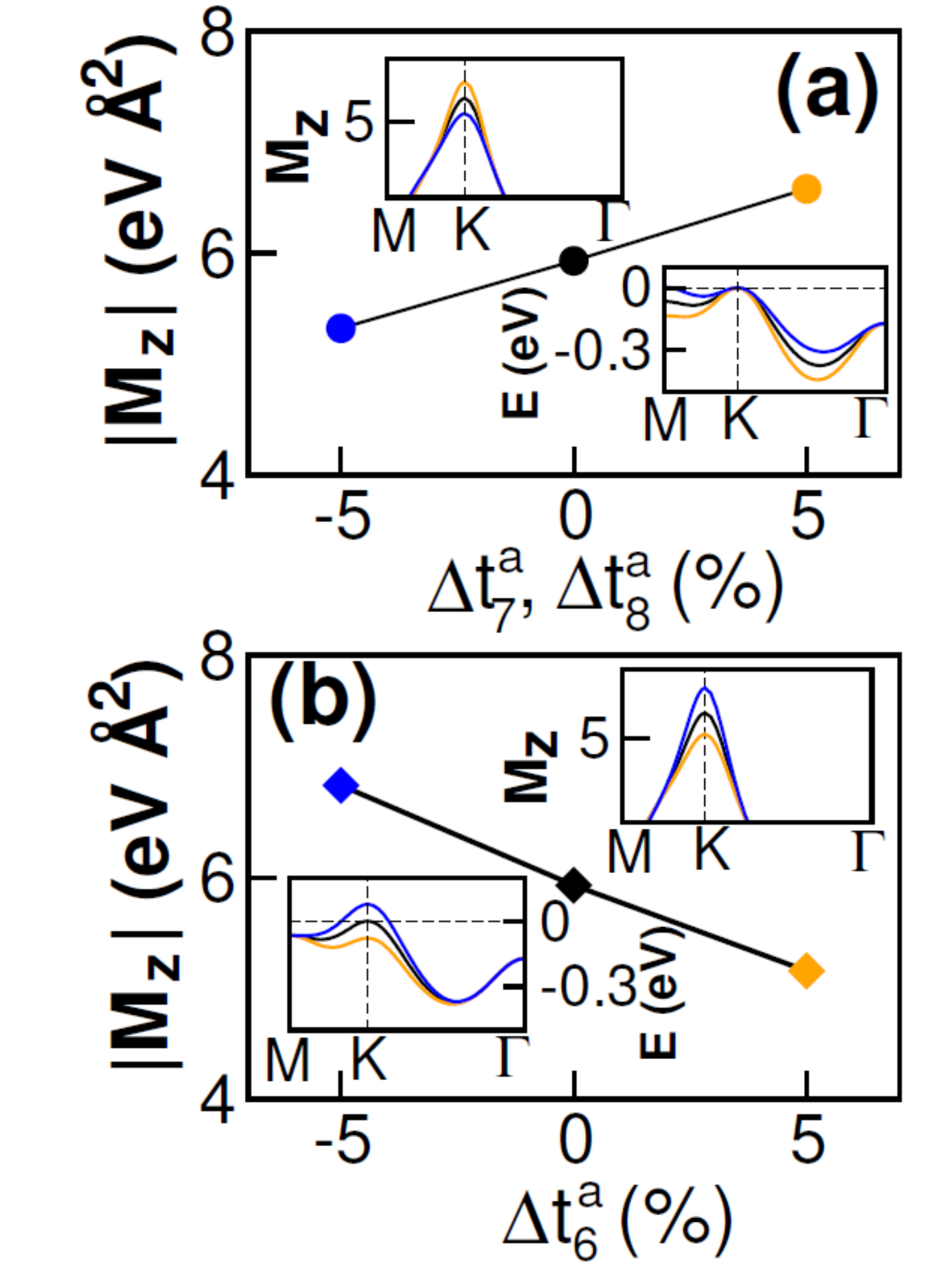}
\caption {Tuning of orbital moment by modifying the appropriate hopping parameters. (a) Variation of the 
 orbital moment at the valley points $K$, $K'$ with the hopping parameters $t_7^a$ and $t_8^a$, with both changed simultaneously, while 
 keeping the other TB parameters in Table \ref{tab1} fixed. Top inset shows the orbital moments near $K$, while the bottom inset shows
 the 
variation of the band structures for the three different cases: $\Delta t =0 \%$ (black lines), 
$\Delta t = +5  \%$ (orange lines), and $\Delta t =- 5 \%$ (blue lines).
(b) The same as in (a) for the hopping parameter $t_6^a$. The hopping parameters in (a) change the valence band curvature, while in (b), $\Delta t_6^a$ shifts the valence band top, thereby changing the gap value $\Delta$
at the valley points.
}
\label{fig4b} 
\end{figure}

The orbital moment for the valence band of the Hamiltonian (\ref{H}), computed using Eq. (\ref{orb}),  is shown in Fig. \ref{fig2} (b) along  selected symmetry lines. 
We find that $\vec M (\vec k)$  is dominated by the first term in Eq. (\ref{orb}), viz.,
the  self-rotation contribution of the wave packet. 
As seen from Fig. \ref{fig2} (b), the orbital moment $ M_z(\vec k)$ has a maximum value at the valley points, while away from the two valleys it decreases and eventually becomes zero at the $\Gamma$ point. 
Furthermore, the direction of $M_z(\vec k)$ is opposite at the $K$  and $K'$ points as dictated by symmetry, leading to what we call the {\it valley-orbital locking}. This is akin to the ``valley-spin locking", commonly discussed in the TMDCs. However, it is interesting to note that while valley-spin locking appears only in the presence of the SOC, the existence of the ``valley-orbital locking'' does not require any SOC (the effect of SOC is not considered in the Hamiltonian \ref{H}), but rather that it originates from the broken inversion symmetry of the crystal. Furthermore, in  presence of the SOC, the ``valley-orbital locking'' gives rise to the ``valley-spin locking" due to the $\lambda \vec L \cdot \vec S$ spin-orbit interaction term.
 
 The orbital moment and its opposite directions at the two valley points can be well understood from the eigenstate $|v \rangle = (\sqrt{2})^{-1} (|x^2-y^2 \rangle + i\tau |xy \rangle)$ at the $K$  and $K'$ points, which carries an orbital angular momentum $l_z = 2\hbar$ for the $K$ point ($\tau = + 1$)  and $l_z = -2\hbar$ for the $K^\prime$ point ($\tau = - 1$) respectively. Recalling the fact that such hybridized orbitals resulted from the inter-orbital hoppings which are induced by the broken $\mathcal{I}$ symmetry, it easily follows that the orbital moment $\vec M(\vec k)$ is a consequence of the lack of $\mathcal{I}$ symmetry. Similarly, the absence of orbital moment at the $\Gamma$ point can be understood from the corresponding eigenstate  $|3z^2-r^2 \rangle$ that carries a zero orbital angular momentum ($l_z = 0$).

 {\it Tuning of the orbital moment} -- 
 Certain hopping parameters can have a large effect on the orbital moments, which we discuss now.
  This provides a tool to manipulate the valley orbital moments, and eventually the OHE, by band structure engineering, which can be achieved by applying strain \cite{Peelaers}, for example.   
 
Diagonalization of the Hamiltonian (\ref{H}) shows that the 
 three orbitals of interest near the $K, K'$ points are  $x^2-y^2, \ xy,$ and $3z^2-r^2$, which constitute the valence and conduction bands [see Fig. \ref{fig2} (a)]. We develop a minimal four-band model later in Section \ref{4band}, but it is clear that TB parameters involving these states, viz., $t_6^a$, $t_7^a$, and $t_8^a$, will have the largest effect on controlling the orbital moments. 
 In fact, as we will see later from the effective four-band model, $t_6^a$ controls the band gap $\Delta$,
 while the other two parameters control the band curvatures (band mass) of the valence band near the valley points. This is also seen from the lower insets in Figs. \ref{fig4b} (a) and (b). 
 As the band curvature increases, the orbital moment increases, while
 as the band gap $\Delta$ increases along with $t_6^a$, the orbital moment changes in the opposite direction. 
 Later we will see from the four-band model that the valley point moments scale as $m_0 \sim t^2/\Delta$, 
 where the hopping term ($t_7^a$, and $t_8^a$ entering in $t$) controls the band curvature
 and $\Delta$ is the band gap. 
 This dependence  is consistent with the full TB calculations presented in Figs. \ref{fig4b}  (a) and (b), as it should simply because the four-band model is extracted from the full TB Hamiltonian Eq. (\ref{H}).

 \begin{figure}[t]
\centering
\includegraphics[width=\columnwidth]{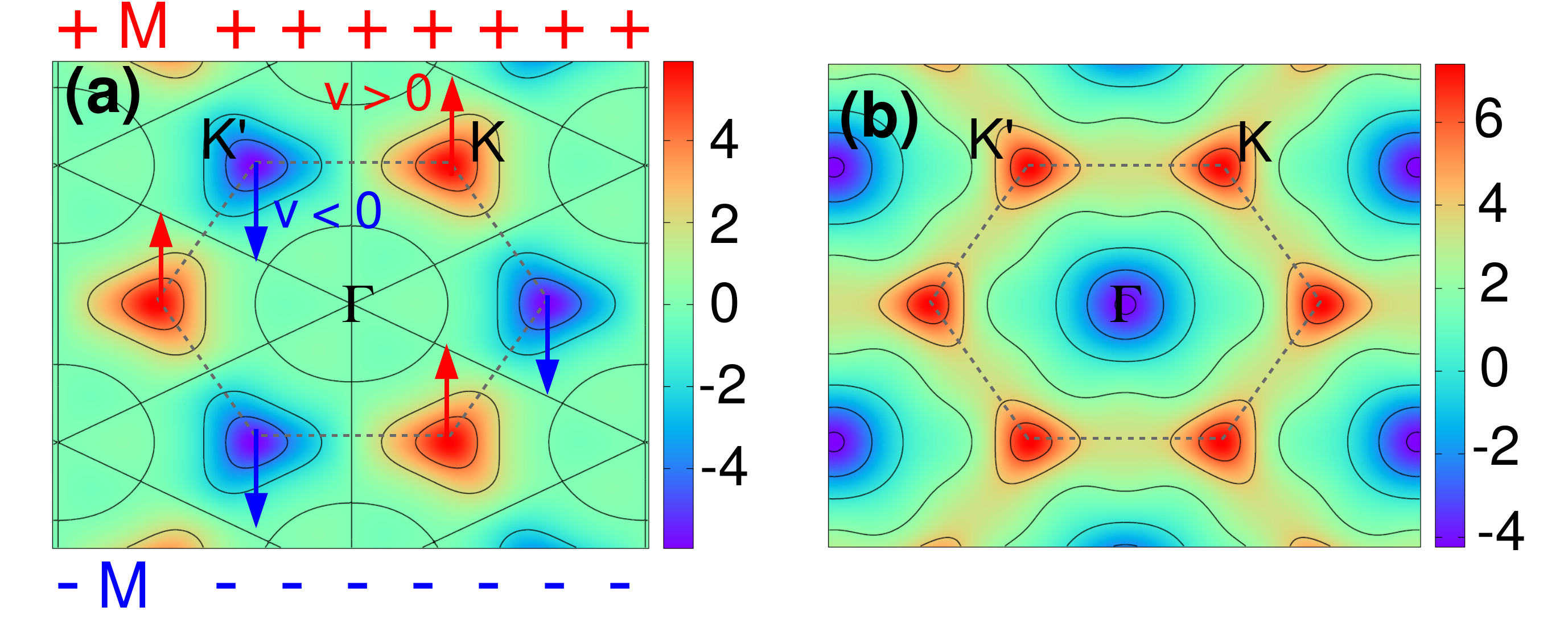}
\caption {Intrinsic orbital moment (in units of eV$\cdot$\AA$^2$) (a) and the orbital Berry curvature (in units of \AA$^2$) (b), on the $k_x$-$k_y$ plane for the lowest occupied band in Fig. \ref{fig2}. Hexagons indicate the Brillouin zone. Results are obtained using the Eqs. (\ref{orb}) and (\ref{obc}) for the TB Hamiltonian (\ref{H}) with the parameters, listed in Table \ref{tab1}. The arrows in (a) indicate the directions of the anomalous velocities, which is opposite for the $K$, $K'$ valleys due to the opposite sign of the Berry curvatures. This leads to an orbital Hall current. The $\pm$ sign in (a) shows the accumulation of the opposite orbital moments at the opposite sides of the sample, resulting from the orbital current flow.  
}
\label{fig3} 
\end{figure}

 Another means of modifying the orbital moment is to just shift the Fermi energy (E$_{\rm F}$) by hole doping, so that less number of bands are occupied. Recently there have been reports of successful hole doping in the TMDCs \cite{Expt-hole}. 
 Since the doped holes will occupy the valley points
 $K$ and $K'$, where the orbital moment is the largest, and these electron states would not participate in transport, hole doping is expected to substantially change the OHE, as we discuss later.
 

\subsection{Orbital Hall Effect } 

 The orbital Hall conductivity (OHC) can be computed from the the momentum sum of the orbital Berry curvatures \cite{Go,Jo}
\begin{equation}\label{OHC} 
  \sigma^{\gamma,\rm orb}_{\alpha \beta}   =  -\frac{e} { N_k V_c} \sum_{n \vec  k}^{occ} \Omega^{\gamma,\rm orb}_{n,\alpha \beta} ({\vec  k}),
\end{equation}
where $\sigma^{\gamma,\rm orb}_{\alpha\beta}$ is given by  
$j^{\gamma, \rm orb}_\alpha = \sigma^{\gamma,\rm orb}_{\alpha\beta} E_\beta$,
$j^{\rm orb,\gamma}_\alpha $ being the orbital current density
   along the $\alpha$ direction corresponding to the $\gamma$ component of the
orbital moment, generated by the  electric field pointed along the $\beta$ direction.
  In the 2D systems, $V_c$ corresponds to the surface unit cell area, so that
  the conductivity has the dimensions of $(\hbar / e)$ Ohm$^{-1}$. Note that the symmetry of the structure dictates that there is only one independent component of the OHC, viz.,   $\sigma^{z,\rm orb}_{yx}$ = -$\sigma^{z,\rm orb}_{xy}$.

  The orbital Berry curvature $\Omega^{\gamma,\rm orb}_{n,\alpha\beta}$ in Eq. (\ref{OHC}) can be evaluated using the Kubo formula from the eigenvalues $\varepsilon_{n  \vec k}$ and the eigenfunctions $u_{n{\vec  k}}$, where $n$ is the band index. The expression is
\begin{equation} \label{obc}         
 \Omega^{\gamma,\rm orb}_{n,\alpha\beta} ({\vec  k}) = 2 \hbar   \sum_{n^\prime \neq n} \frac {{\rm Im}[ \langle u_{n{\vec  k}} | \mathcal{J}^{\gamma,\rm orb}_\alpha | u_{n^\prime{\vec  k}} \rangle  
                        \langle u_{n^\prime{\vec  k}} | v_\beta | u_{n{\vec  k}} \rangle]} 
                        {(\varepsilon_{n^\prime \vec k}-\varepsilon_{n  \vec k})^2},
\end{equation}
 where the orbital current operator $\mathcal{J}^{\gamma,\rm orb}_\alpha = \frac{1}{2} \{v_\alpha, L_\gamma \}$,  
with $L_\gamma$ being the orbital angular momentum operator and $v_{\alpha} =  \frac{1}{\hbar} \frac{\partial H }{ \partial k_\alpha}$
is the velocity operator.

 The computed orbital Berry curvature along with the orbital moment over the entire BZ are shown in Fig. \ref{fig3}. As seen from this plot the dominant contribution to  
 the orbital Berry curvature comes from the region close to the valley points. Furthermore, the contributions from the two valleys appear with the same sign which add up to give a net large OHC. It is also interesting to note that although the orbital moment vanishes at the $\Gamma$ point, the orbital Berry curvature is nonzero there and appears with an opposite sign than that near the valley points. However, the magnitude of the contribution of the $\Gamma$ point being much smaller, the net OHC, the  BZ sum of $\Omega^{z,\rm orb}_{yx}$ over the occupied bands, is still dominated by the valley point contributions. 
 The magnitude of the OHC is quite large: $\sigma^{z,\rm orb}_{xy}  
 \approx -10.8 \times 10^3 (\hbar / e ) \Omega^{-1}$ taking a nominal lattice constant $a =3.19$ \AA.

 The above results for the orbital Hall conductivity can be understood by considering the orbital moments and
 their velocities in the momentum space under the applied electric field.  As well known in the literature, the $K, K^\prime $ valley points have Berry curvatures with opposite signs \cite{Xiao}. As a result, under an applied electric field, the electrons in the two valleys acquire an oppositely directed transverse velocity due to the Berry curvature ($\Omega (\vec k)$) term in the semi-classical expression \cite{Niu} 
$ \dot  {\vec  r}_c = \hbar^{-1}[\vec \nabla_k \varepsilon_{k} +e \vec E \times \vec \Omega (\vec k)]_{\vec k_c}  $.
Since the electrons at the two valleys carry  
orbital moments in opposite directions as well, they do not cancel, but add up to produce a net
orbital Hall current, as indicated schematically in Fig. \ref{fig3} (a). 

In contrast to the valley points, the $\Gamma$ point does not have any intrinsic orbital moment. In this case, however, the applied electric field can induce a momentum-space orbital texture by dynamically inducing inter-band superposition as discussed for the centrosymmetric materials \cite{Go,Jo}. This electric field induced orbital texture, in turn, gives rise to a contribution to the orbital Hall conductvity, though with a smaller magnitude than the contributions from the valley points.

Note that the orbital Berry curvature follows a symmetry relationship different from the orbital moment $\vec M (\vec k)$, discussed already. Presence of $\cal T$ symmetry demands that $\Omega^{z,\rm orb}_{yx} (\vec k) = \Omega^{z,\rm orb}_{yx} (-\vec k)$, which is followed by the distribution of $\Omega^{z,\rm orb}_{yx} (\vec k)$ in Fig. \ref{fig3} (b), showing that the contributions from $K$ and $K^\prime$ have the same sign.

{\it Hole doping} -- In view of the fact that the valley points make a major contribution to the OHC, 
it is  clear that hole doping, which depletes the occupation of the valley point states, would considerably diminish the magnitude of the OHC. We have calculated that a small change in the Fermi energy due to hole doping 
($\Delta {E_F} = - 0.05$  eV) leads to a considerable change ($14\%$) in the OHC, 
$\sigma^{z,\rm orb}_{xy}  
 \approx -9.3 \times  10^3 (\hbar / e ) \Omega^{-1}$,
 providing a potential tool to manipulate the OHC in the TMDCs. 

\section{Spin-orbit coupling:  Orbital and Spin Hall effects}     \label{soc}

It is an important point  to note that the orbital Hall effect exists even without the spin-orbit coupling (SOC). Indeed, in the above tight-binding model, there was no SOC term present. The presence of the SOC term will modify the magnitude of the OHE, but only by a small amount. 
The spin Hall effect, on the other hand, exists only due to the SOC, fundamentally because conductivity is measured in real space, and the spin space does not couple to it without the SOC term. In this Section, we
elaborate these points from quantitative calculations.

\begin{figure}[ht!]
\centering
\includegraphics[width=\columnwidth]{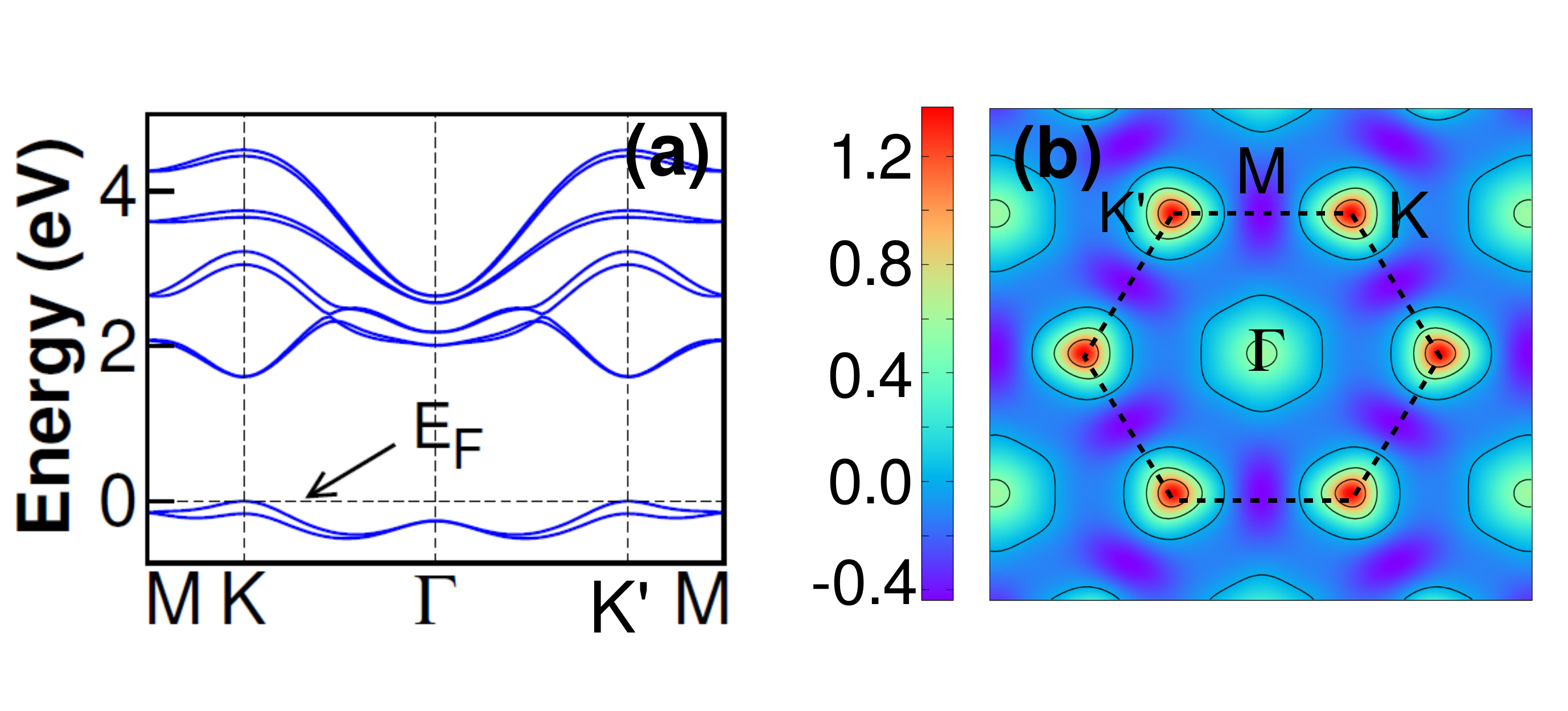}
\caption {
Band structure and spin Berry curvature in presence of spin-orbit coupling. (a) The TB band structure with the SOC parameter $\lambda = 0.08$ eV. (b) The calculated spin Berry curvature (units of \AA$^2$) in the $k_z = 0$ plane, summed over the occupied bands in (a) (insulating case, $n_e = 2$). }
\label{fig4} 
\end{figure}
\begin{figure}[t]
\centering
\includegraphics[width=\columnwidth]{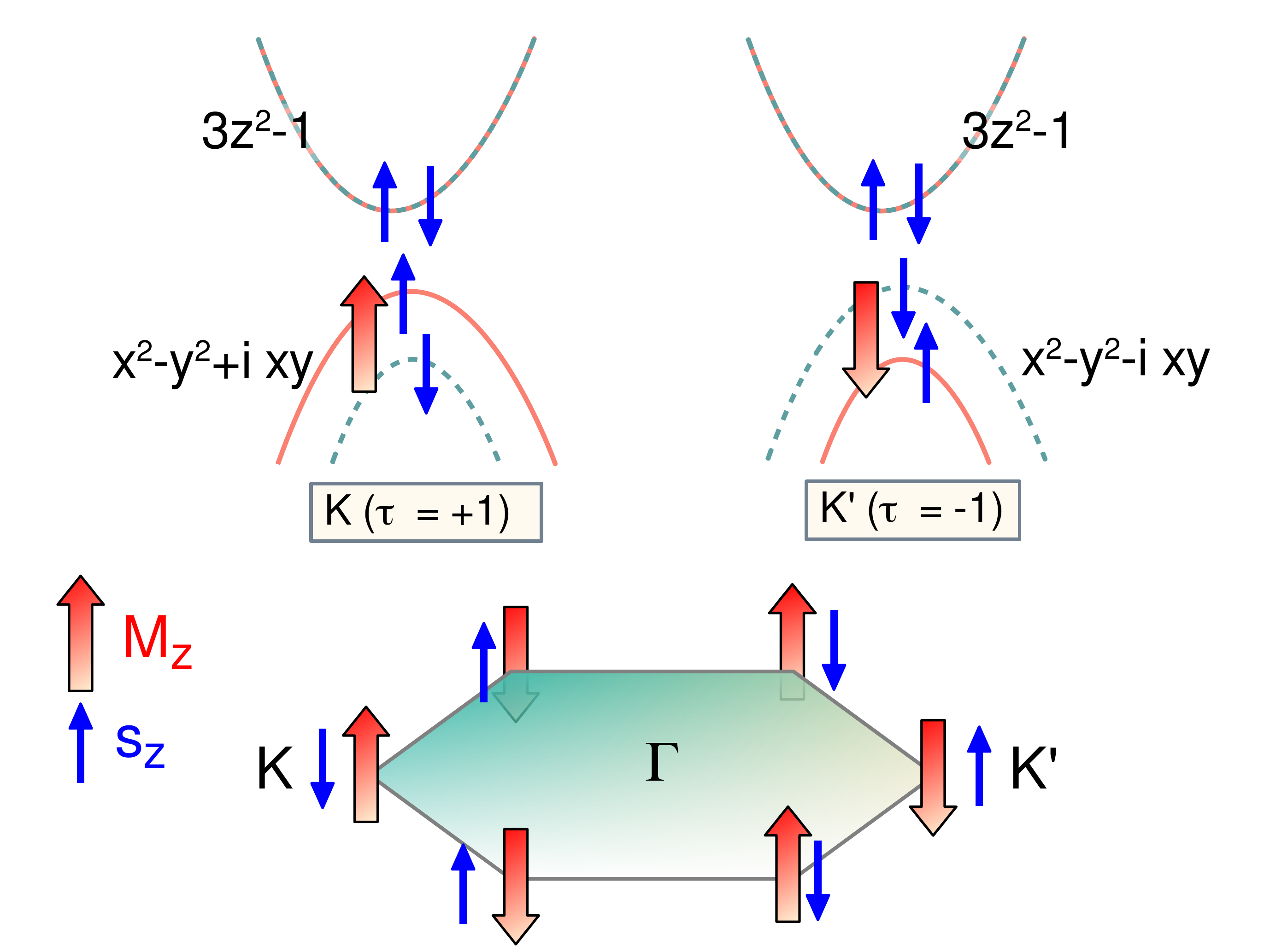}
\caption {Schematic illustration of the valley-orbital  and  valley-spin locking    of the valence bands in the TMDCs. The complex orbitals ``$x^2-y^2 \pm i ~xy$"  lead to the opposite alignment of the orbital moment at the two valleys as shown, which in turn leads to the opposite spin alignment of the valence bands for the two valleys (valley-spin locking)  due to the $\lambda \vec L \cdot \vec S$ term.
}
\label{fig5} 
\end{figure}

The effect of SOC is included by adding a term $ {\cal H}_{\rm SOC} = \lambda {\vec L} \cdot {\vec S}$ to the TB Hamiltonian (\ref{H}). Written in the basis with spin-orbital order: $( \phi_{\alpha \sigma} = 
 xy \uparrow, \ x^2-y^2\uparrow, \ 3z^2-r^2\uparrow, \ xz\uparrow, \ yz\uparrow, \ xy \downarrow, \ x^2-y^2\downarrow, \ 3z^2-r^2\downarrow, \ xz\downarrow,$ and $yz\downarrow )$, it reads
\begin{widetext}
\begin{eqnarray}\label{H_{SOC}}
{\cal H}_{\rm SOC} = \lambda {\vec L} \cdot {\vec S} =
\frac{\lambda}{2}
\left[ 
{\begin{array}{*{20}c}
   0  & \ 2i    & \ 0   & \ 0 & \ 0 & \ 0 & \ 0 & \  0 \ & -i \ & 1\\
   -2i & \ 0    & \ 0   & \ 0 & \ 0 & \ 0 & \ 0 & \ 0 \ & 1 & i \\
   0  & \ 0    & \ 0   & \ 0 & \ 0 & \ 0 & \ 0 & \ 0 & \ -\sqrt{3} \ & \sqrt{3} i \\
   0  & \ 0    & \ 0   & \ 0 & \ -i & \ i & \ -1 & \ \sqrt{3} & \ 0 & \ 0 \\
   0  & \ 0    & \ 0  & \ i & \ 0 & \ -1 & \ -i & \ -\sqrt{3} i & \ 0 & \ 0 \\
   0  & \ 0    & \ 0   & \ -i & \ -1 & \ 0 & \ -2i  & \ 0 & \ 0 & \ 0 \\
   0  & \ 0    & \ 0   & \ -1 & \ i & \ 2i & \ 0  & \ 0 & \ 0 & \ 0 \\
   0  & \ 0    & \ 0   & \ \sqrt{3} & \ \sqrt{3}i & \ 0 & \ 0  & \ 0 & \ 0 & \ 0 \\
   i  & \ 1 \ & -\sqrt{3} \ & 0 & \ 0 & \ 0 & \ 0 & \ 0 & \ 0 & \ i \\
   1 & \ -i & \ -\sqrt{3}i & \ 0 & \ 0 & \ 0 & \ 0 & \ 0 & \ -i & \ 0 \\
\end{array} }  \right].
\end{eqnarray}
\end{widetext}
The resulting band structure with the spin-orbit coupling included 
is shown in Fig. \ref{fig4} (a)
 for a typical strength of the SOC parameter. 
The key new feature of the band structure is the spin splitting of the valence band  with opposite spin moments at the two valleys. 
This leads to the so called ``valley-spin coupling" \cite{Xiao}, where the valence band top 
at one valley has a positive spin moment, while the other has a negative spin moment, coupling thereby  the spin and valley degrees of freedom (see Fig. \ref{fig5}). 
The valley dependent spin splitting directly follows from the computed orbital moment ($M_z$) at the $K$, $K^\prime$ points and the $\lambda \vec L \cdot \vec S$ SOC term, which splits the previously spin-degenerate valence band into two, with the spin of the lower energy state (bottom valence band) at a particular valley aligned oppositely to the  orbital moment at that valley. 
The spin-$\downarrow$ band is lower in energy at the $K$ valley ($\tau = 1$), while the spin-$\uparrow$ band is lower in energy at the $K^\prime$ valley ($\tau = -1$), with a spin splitting of $\approx 2 \lambda$.

The spin Hall conductivity can be calculated by computing the BZ sum of the
spin Berry curvature $\Omega^{z,\rm spin}_{\nu, yx} (\vec k)$, evaluated using  the 
band energies and the eigen functions. 
The expression for the  spin Berry curvature 
is analogous to Eq. (\ref{obc}) for the orbital Berry curvature,
 where the orbital current operator $\mathcal{J}^{\gamma,\rm orb}_\alpha$
  should be  replaced 
by the spin current operator $\mathcal{J}^{\gamma,\rm spin}_\alpha = \frac{1}{4} \{v_\alpha, s_\gamma \}$,
 $s_\gamma$ being  the Pauli matrices for the electron spin. 
 Note that the spin $\uparrow$ and  $\downarrow$ bands have opposite contributions to the spin Berry curvature, although their magnitudes are not exactly the same due to the broken $\mathcal I$ symmetry. 
 However, they nearly cancel each other
 at each momentum point, if both spin bands are occupied, e.g., for the insulating Mo and W compounds, which results in a net small contribution to the SHC as seen from Fig. \ref{fig4} (b).
 
 The analytical expressions for these contributions are discussed further in section \ref{4band}
 within the four-band model. 
It turns out that in Fig. \ref{fig4} (b), 
the contributions from the $M$ points in the BZ  dominate over the valley point contributions, 
leading to a small SHC of the magnitude
$\sigma^{z,\rm spin}_{yx} \approx 2 (\frac{\hbar}{e}) \ \Omega^{-1} $.


%
 \begin{figure}[t]
\centering
\includegraphics[scale=0.25]{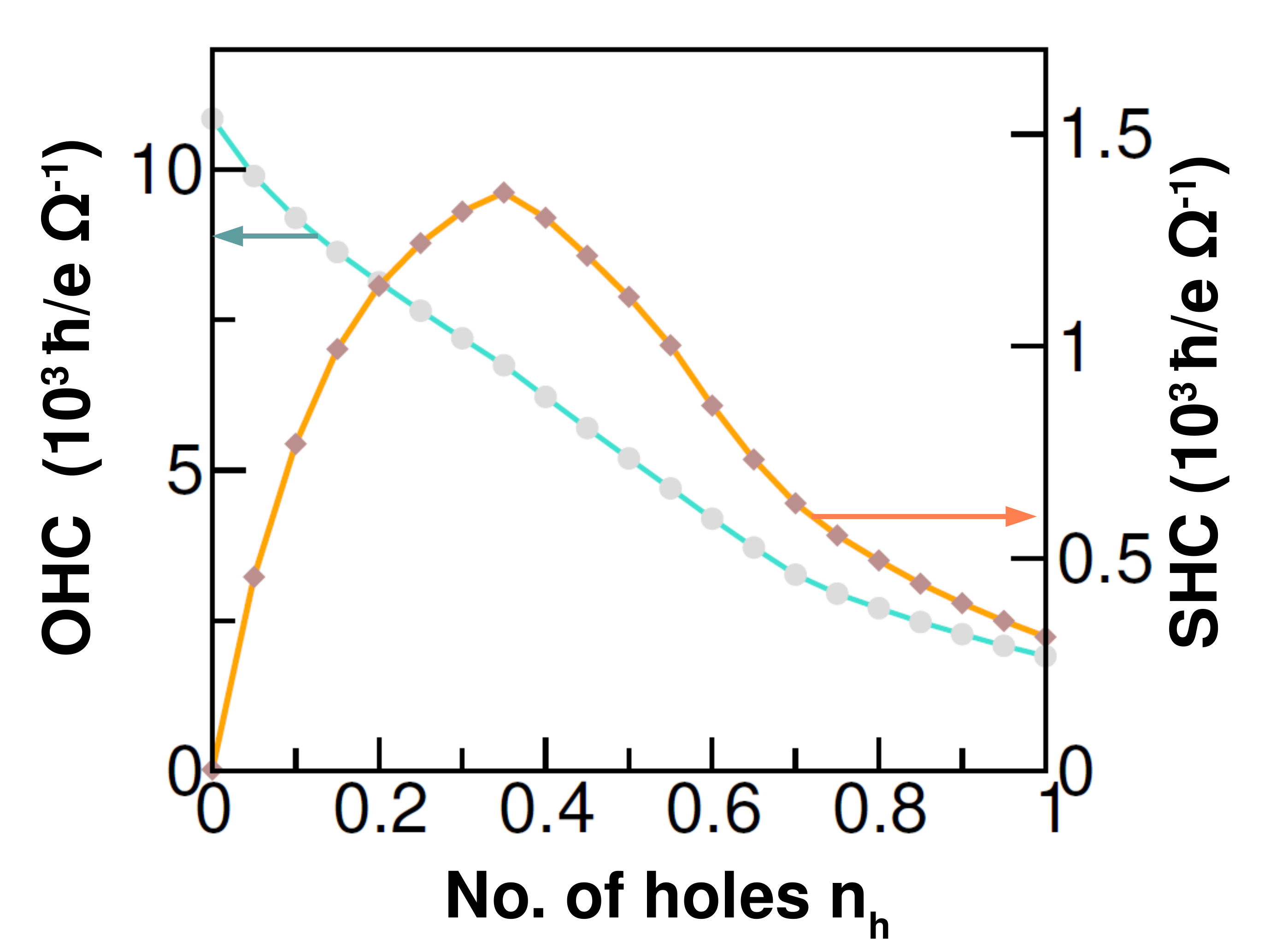}
\caption { Effect of hole doping on OHC ($- \sigma^{z,\rm orb}_{yx}) $ and SHC ($\sigma^{z,\rm spin}_{yx} $).
Zero on the x-axis corresponds to the insulating state with   $n_e = 2$. Here  $\lambda = 0.1$ eV. 
}
\label{fig_ohc} 
\end{figure}
%

 \begin{figure}[ht!]
\centering
\includegraphics[width=\columnwidth]{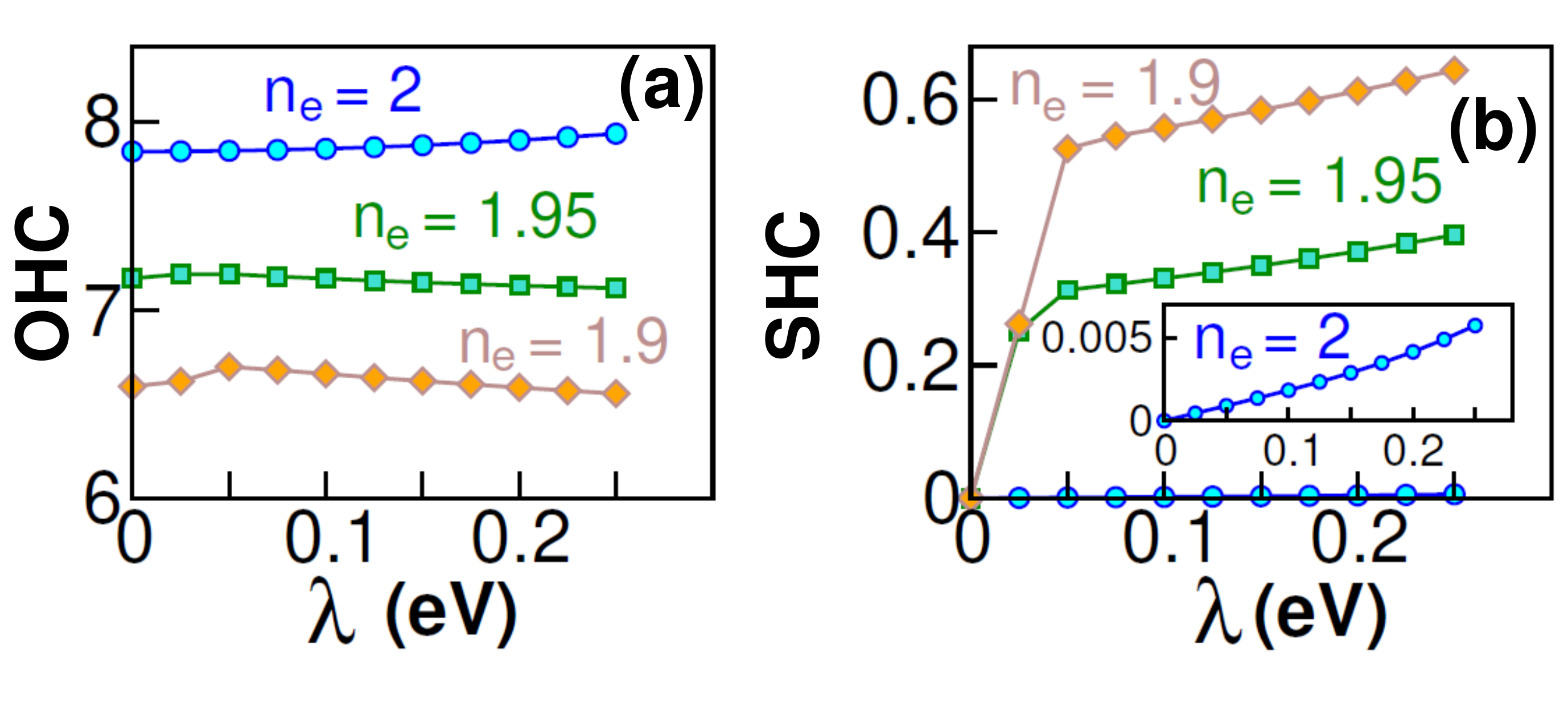}
\caption {
Effect of the spin-orbit coupling on the OHC and SHC, studied within the TB model. 
(a) The variation of the OHC (-$\sigma^{z,\rm orb}_{yx}$) with $\lambda$ for three different occupations ($n_e$) of the $d$ orbital. 
Note that the occupancy $n_e = 2 $ corresponds to the valence bands fully occupied
with the Fermi energy $E_F$ as indicated in Fig. \ref{fig4}(a). 
(b) The same for the SHC $\sigma^{z,\rm spin}_{yx}$ with the same color coding as in (c).
The inset shows the expanded view of SHC for $n_e = 2$, showing the near about linear dependence on $\lambda$. 
The conductivities in (a) and (b) on the y-axes are in units of $1.405 \ a^{-2}  \times 10^4 (\frac{\hbar}{e}) \Omega^{-1}$,
where $a$ is the lattice constant.
Note that the magnitudes of the conductivities are significantly changed by changing the electron
occupancy, leading to the tuning of the effects via hole doping.
}
\label{fig7b} 
\end{figure}

Note that the magnitude of the SHC is three orders of magnitude smaller than the corresponding value of the OHC. This is advantageous from the practical point of view as it overcomes the common difficulty in detecting the orbital contribution to the angular momentum Hall current \cite{Go2019}.
Our calculation suggests that the TMDCs, which have been widely studied for the SHE \cite{Xiao,Feng,Shen}, have a rather strong intrinsic OHE with possible applications in orbitronics.

{\it Hole doping and effect of spin-orbit coupling:}
TMDCs also provide an excellent platform to manipulate the OHE and the SHE, e.g., by hole doping. 
Reducing the  occupancy  of the $d$ orbitals 
makes the spin-polarized bands only partially occupied, which
 leads to a drastic enhancement of the SHE, 
since the spin-split bands at individual momentum points no longer nearly cancel any more. 
As seen from Fig. \ref{fig_ohc}, the SHC can be as large as $\approx 10^3 (\hbar/e) \Omega^{-1}$ for an electron occupancy $n_e \approx 1.6$ (number of holes $n_h = 2- n_e \approx 0.4 $), as
compared to $\approx 2 (\hbar/e) \Omega^{-1}$ for $n_e \approx 2$.
It is important to note that SHC does not increase monotonically with the hole doping and beyond a certain hole concentration it starts to fall off. Obviously, for $n_h = 2$, i. e., for completely empty bands in our model, SHC must go to zero as there are no occupied electron states to contribute to the conductivity any more.
The SHC is still larger for single hole doping ($n_h = 1 $) compared to the undoped case, which suggests the presence of a larger SHE in metallic TMDCs, such as NbS$_2$ with a single valence $d$ electron,
a result that is confirmed from the density-functional results presented later.

In contrast to the SHE, OHE decreases with hole doping.  This is because the shift in the Fermi energy due to hole doping removes electrons from the valley points that had dominant contribution to OHE
thereby drastically decreasing the magnitude of the effect (see Fig. \ref{fig_ohc}). 
Thus with  hole doping, it is also possible to control the ratio between OHC and SHC, i.e., the orbital and the spin contribution to the net angular momentum current, which may be of practical importance. 
We also note that the orbital and spin contributions have opposite signs, i. e., at a particular edge of the sample, the direction of the accumulated orbital moments is opposite to that of the spin moments. 
This is, again, due to the $ \lambda  \vec L \cdot \vec S $ term in the Hamiltonian
 that favors the anti-parallel alignment of spin and orbital moments in TMDCs.

The effect of the SOC parameter $\lambda$ on the OHE and SHE are  shown in 
Fig. \ref{fig7b}. 
As already mentioned,
the OHC exists even for $\lambda = 0$, while the SHC necessarily needs the presence of the SOC.
In this sense, the OHE may be considered more fundamental than the SHE.
While the dependence of OHE on $\lambda$ is relatively weak as seen from Fig. \ref{fig4}, 
the SHC changes drastically with the strength of the SOC, increasing with increasing strength of $\lambda$.
Similar behavior is also reported for the centrosymmetric systems \cite{Jo}.

The key results of the NN TB model may be summarized as follows. 
(1)  The monolayer TMDCs host an intrinsic orbital moment, which can be controlled by manipulating the band curvature and the band gap near the valley points, by applying strain for example.
(2) The valley orbital moment, the OHE, and the SHE, all can be controlled by hole doping. 
(3)  We studied the effect on both OHC and SHC, when the SOC is included. 
While OHC exists even without the SOC, the SHC requires a non-zero $\lambda $, which also
determines the relative magnitudes of the two. In this sense the OHE is more fundamental than the SHE, since the former
can exist with or without the SOC.


{\it Orbital Hall effect and valley Hall effect} --
The orbital Hall effect is distinct from the well-studied valley Hall effect (VHE)
\cite{Xiao,Xiao_prl,ScRep}, though both involve motion of the electrons in the
two different valleys in opposite directions. 
%
 While the VHE refers to the accumulation of the charge carriers with opposite valley indices at the opposite edges, OHE corresponds to the accumulation of the opposite orbital moments at the edges of the sample.
As a result OHE is a property of the intrinsic material, 
while the VHE is an extrinsic effect, in the sense that  in order to detect the VHE,
a population imbalance between the two valleys must be  created by some extrinsic means
such as 
light illumination.
 Furthermore, by creating such a valley population imbalance, the time reversal  symmetry  is explicitly broken, leading to a net orbital magnetization in the system. 
 In contrast, detection of OHE does not require any such population imbalance, and the time-reversal symmetry of the system is kept in tact, resulting in vanishing  orbital magnetization in the system.

%




\section{Effective four-band model}\label{4band}

In this Section, we construct a minimal four-band TB model near the valley points, $K$ or $K'$, 
since the physics of the OHE in the TMDCs is dominated by the valley point contributions. 
The model is meant to describe the low-lying states of the four bands, including spin, at the valley points as sketched
in Fig. \ref{fig5}.
Our earlier work for the undoped case \cite{ohe_our} is generalized here for the  case with hole doping. 
Analytical results for the OHE and the SHE are obtained, which provide considerable insight 
into the physics of the problem.



\subsection{The four-band model} \label{4band}
We begin our discussion with the  construction of the effective four-band  TB model.  
As discussed in Section \ref{orbandohe}, the $|xz \rangle$ and $|yz \rangle$ orbitals are decoupled from the lower lying $|xy \rangle,|x^2-y^2 \rangle$, and $|3z^2-r^2\rangle$ orbitals around the valley points.  We, therefore, expand the Hamiltonian (\ref{H}) around the $K (-4\pi/3, 0)$ and $K^\prime (4\pi/3, 0)$ points of the BZ, keeping only these three orbitals in the basis set and keep only terms of the order of $q$, where $q \equiv k -K$ or $k-K'$. The resulting Hamiltonian has the following form in the basis $\{|xy \rangle,|x^2-y^2 \rangle$, $|3z^2-r^2\rangle\}$:

\begin{eqnarray}     \label{3band}  
{\cal H} (\vec q ) &= 
\left[
{\begin{array}{*{20}c}
    h^q_{11} & h^q_{12} & h^q_{13}  \\
    (h^q_{12})^* & h^q_{22} & h^q_{23}  \\
    (h^q_{13})^* & (h^q_{23})^* & h^q_{33} \\
  \end{array} }  \right],
\end{eqnarray} 
where
\begin{eqnarray} \nonumber
 h^q_{11} &=& \varepsilon_1-(t^a_1+2t^b_1)+ \tau \sqrt{3}q_xa (t^b_1- t^a_1)\\ \nonumber
 h^q_{22} &=& \varepsilon_1-(t^a_2+2t^b_2)+ \tau \sqrt{3}q_xa (t^b_2- t^a_2)\\ \nonumber
 h^q_{33} &=& -3t^a_3\\ \nonumber
  h^q_{12} &=& \tau i \sqrt{3} (t^a_6 +t^b_6+t^c_6) + \frac{3 \tau q_ya}{2} (t^b_6-t^c_6)  \\ \nonumber
 h^q_{13}&=&  i \frac{q_xa}{2}(-2t^a_7 +t^b_7+t^c_7) + \frac{ 3 \tau q_ya}{2} (t^b_7-t^c_7) \\ \nonumber
 h^q_{23}&=& \frac{ \sqrt{3} \tau q_xa}{2} (-2t^a_8 +t^b_8+t^c_8) + i\frac{ \sqrt{3} q_ya}{2} (t^b_8-t^c_8). 
\end{eqnarray}
Here, $\tau = \pm 1$ is the valley index
for the $K$ and $K^\prime$ valleys respectively. 
As discussed in Section \ref{orbandohe}, $|u \rangle = (\sqrt{2})^{-1} (|x^2-y^2 \rangle + i\tau |xy \rangle)$ and $|d\rangle=|3z^2-r^2\rangle$ constitute the valence band and the conduction band respectively near the two $K, K^\prime$ valleys. 
Therefore, we, further, transform the Hamiltonian (\ref{3band}) in the   
  orbital pseudo-spin basis set $ \{|u \rangle, |d\rangle \}$, which then takes the simple form,
\begin{eqnarray} \nonumber    \label{HK}  
{\cal H}_v (\vec q ) = 
\left[
{\begin{array}{*{20}c}
    -\Delta/2 & \tau tq_xa+ itq_ya   \\
    \tau tq_xa- itq_ya  & \Delta/2\\
   \end{array} }  \right]
   = \vec d \cdot \vec \sigma ,\\
    \end{eqnarray} 
which is valid in the regions near the valley points, $K$ and $K'$, in the Brillouin zone.
Here $\sigma_x, \sigma_y$, and $\sigma_z$ are the Pauli matrices for the orbital pseudo-spins, $d_x = \tau tq_xa, d_y = -tq_ya,$ and $ d_z=-\Delta/2$, where $a$  is the lattice constant.
 We note that the Hamiltonian ${\cal H}_v (\vec q )$ is similar to the Hamiltonian for graphene  except for the onsite mass term, representing a massive Dirac particle, which is relevant for the gapped graphene system \cite{Xiao_prl}.

There are just  two parameters in this effective Hamiltonian (\ref{HK}), 
viz., the energy gap parameter $\Delta$ and 
a generalized hopping parameter $t$, which can be expressed in terms of the $d$-$d$ hopping parameters discussed in Section \ref{alld},
viz.,
\begin{eqnarray} \label{TB-parameters}
\Delta &=& (t_1^a + 2t_1^b) -\epsilon_1 -\sqrt{3}(t_6^a+t_6^b+t_6^c)- 3t_3^a  \nonumber \\
&=& 3(t_1^a + t_2^a)/2 -
 \epsilon_1 - 3\sqrt{3}t_6^a - 3t_3^a , \nonumber \\
t &=& (\sqrt{3}/2\sqrt{2}) [(t_8^c-t_8^b)-\sqrt{3}(t_7^c -t_7^b)] \nonumber \\
&=& 3 (t_7^a + \sqrt{3}t_8^a)/(2\sqrt{2}),
  \end{eqnarray} 
 where the second equalities in both equations are obtained by using the rotational symmetry of the structure.
 For the hopping parameters listed in Table \ref{tab1} and Table \ref{tab3}, the numerical values of these   are: $\Delta = 1.69$ eV and $t = 1.02$ eV.

In order to include the effect of the SOC, we express the Hamiltonian $\cal{H}_{\rm SOC}$,  Eq. (\ref{H_{SOC}}), in the orbital pseudo-spin subspace $\{ |u \uparrow \rangle, |d \uparrow \rangle, |u \downarrow \rangle, |d \downarrow \rangle\}$, with the result  
 \begin{eqnarray} \nonumber   \label{HqSOC}  
{\cal H}^{\rm SOC}  &=& 
\left[
{\begin{array}{*{20}c}
    \tau \lambda& 0 & 0 & 0 \\
    0  & 0 & 0 & 0\\
    0 & 0 &  - \tau \lambda& 0 \\
    0 & 0 & 0  & 0
    \end{array} }  \right], \\
\end{eqnarray} 
 where, again, $\lambda$ is the strength of the SOC.
Adding this term to the Hamiltonian (\ref{HK}), we can construct the following effective four-band TB model in the basis ($ |u \uparrow \rangle, |d \uparrow \rangle, |u \downarrow \rangle, $ and $ |d \downarrow \rangle$):
\begin{widetext}
 \begin{eqnarray} \nonumber   \label{HKSOC}  
{\cal H}_v^{\rm SOC} (\vec q ) &=&  {\cal H}_v(\vec q ) \otimes I_s + 
{\cal H}^{\rm SOC} 
=  (\vec d \cdot \vec \sigma )  \otimes I_s   + \frac{\tau \lambda} {2}  (\sigma_z+1) \otimes s_z \\
&=& \left[
{\begin{array}{*{8}c}
    -\Delta/2 + \tau \lambda& \tau tq_xa+ itq_ya  & 0 & 0 \\ \nonumber
    \tau tq_xa- itq_ya  & \Delta/2 & 0 & 0\\
    0 & 0 & -\Delta/2 - \tau \lambda& \tau tq_xa+ itq_ya \\
    0 & 0 & \tau tq_xa- itq_ya  & \Delta/2
   \end{array} }  \right],\\
 \end{eqnarray} 
 \end{widetext}
where $I_s$ and $\vec s$ are respectively the $2 \times 2$ identity matrix and Pauli matrices in the spin space,
and $\vec \sigma$ is as defined before the orbital pseudo spin. 
  Note that Eq. (\ref{HKSOC}) is consistent with the effective Hamiltonian reported
earlier \cite{Xiao,ohe_our}. However,
the TB derivation has the benefit that it directly expresses the parameters of the Hamiltonian in terms of the TB $d$-$d$ hopping integrals.

\subsection{ Band structure, Orbital moment and Orbital Hall effect}
\label{b}

It is possible to obtain analytical results for the four-band model, which provides considerable insight into
the orbital Hall effect, both in the undoped and the doped cases. 
For the hole doped case, since both OHE and SHE are dominated from the 
contribution from the valley points, for which the four-band model is valid,
both are described quite well in the model. For the undoped case, however,
while this is true for 
the OHE, this is not so for the SHE. 
The reason is that since both spin bands are occupied, 
the SHC contributions from each of the spin polarized bands nearly 
cancel each other, and the effect is
no longer dominated by the valley points in the BZ
[see Figs. (\ref{fig4}) and (\ref{fig7b})].


Diagonalizing the Hamiltonian (\ref{HKSOC}), we get
the spin polarized band structure near the valley points, which is
\begin{equation} \label{evalue}
\varepsilon^\nu_{\pm} (\vec q) =   2^{-1} [\tau \nu  \lambda \pm ((\Delta - \tau \nu  \lambda)^2 + 4 t^2 a^2 q^2)^{1/2}] .
\end{equation} 
Note that by keeping only terms linear in $\vec q$ in the Hamiltonian (\ref{HKSOC}), we have ignored here the higher-order terms causing the trigonal warping of the energy contours. 
We show later in section \ref{warping} that this does not affect the basic physics of the OHE. 


 \begin{figure}[t]
\centering
\includegraphics[width=\columnwidth]{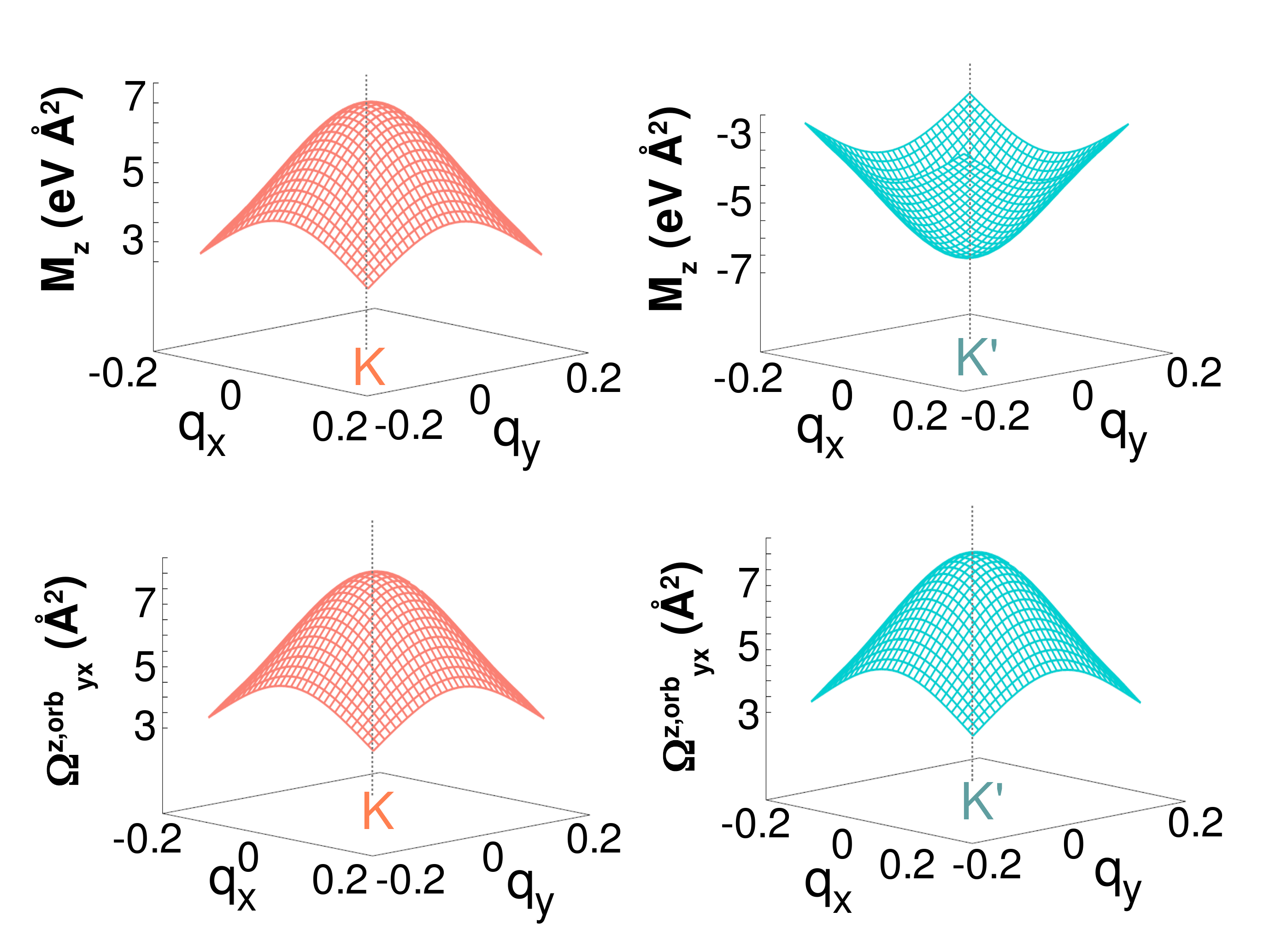}
\caption {Orbital moment (top) and the orbital Berry curvature (bottom) at the valley points, computed from the effective four-band model (\ref{HKSOC})
using Eqs. (\ref{orbk}a) and (\ref{obck}). 
All plots are for the top most valence bands (which is spin $\uparrow$ for the $K$
and spin $\downarrow$ for the $K'$ valleys, $\nu = +1$ and $-1$, respectively).
Note that the orbital moments have opposite signs, while the orbital Berry curvature has the same signs at the 
two valleys.
Hamiltonian parameters are: $t = 1.02 $ eV, $\Delta =$ 1.69 eV, $\lambda = 0.08 $ eV, and $a =$ 3.19 \AA.}
\label{TBfig} 
\end{figure}

{\it Orbital moment} --
The orbital moments can be computed in a straightforward manner from  Eq. (\ref{orb}) from the eigenvalues (\ref{evalue}) and the corresponding eigen functions for the two valence bands. 
The result depends on the valley index ($\tau = \pm 1$) and the spin state
($\nu = \pm 1$ for $\uparrow$ and $\downarrow$ spin respectively) of the band.
A straightforward calculation yields the result
\begin{subequations}     \label{orbk}         
\begin{gather} 
\hspace{-30mm}
M_z (\vec q) =
 \frac{\tau m_0    D^\nu (D^{-\nu} -\lambda)  \Delta }       {2[(D^\nu)^2 + t^2q^2a^2]^{3/2}}\\
 \approx  \tau m_0 [1+\lambda(3 \nu \tau - 2)/\Delta] (1- 6 \  m_0 q^2 / \Delta),
 \end{gather}
\end{subequations}
where $m_0 =  \Delta^{-1} t^2 a^2 $ is the magnitude of the orbital moment at the valley points ($q = 0$) in absence of SOC ($\lambda = 0$) and $ D^\nu = (\Delta  - \nu \tau \lambda)/2 $. The second line is the expansion for small $q$ and $\lambda$, both $t a q$ and $\lambda  \ll \Delta$. 
Eq. (\ref{orbk}) shows that $M_z (\vec q)$ depends on the SOC parameter $\lambda$, but the dependence is weak.
The calculated orbital moment $M_z (\vec q)$ is shown in Fig. \ref{TBfig} near the two  valley points  $K (\tau = +1)$ and  $K' (\tau = -1)$. 
As  is clear from Eq. (\ref{orbk}) as well as Fig. \ref{TBfig}, $M_z (\vec q)$ has opposite signs for the two valleys ($\tau = \pm 1$), representing the valley-orbital coupling as discussed in Section \ref{orbandohe}.

{\it Orbital Hall Conductivity} --
The TMDCs are well known for their valley degrees of freedom, viz., the two valleys have opposite Berry curvatures,
 which results in an oppositely directed anomalous velocity under an applied electric field. 
 As discussed in Section \ref{orbandohe}, this gives rise to an orbital Berry curvature, leading to the OHE. The analytical expression of this orbital Berry curvature for the two valence bands ($\nu = \pm 1$) of the four-band model (\ref{HKSOC}) can be computed using Eq. (\ref{obc}), which yields the result 
\begin{equation}\label{obck}
 \Omega^{z,\rm orb}_{\nu,yx} (\vec q)  
 =  \frac{2\tau M_z (\vec q)}
 { \Delta+    \lambda (\nu \tau -2) }.
\end{equation}
 Here $M_z (\vec q)$ is the orbital moment near the valley points as given in Eq. (\ref{orbk}).  The calculated orbital Berry curvature as well as the orbital moment near the 
 valley points are shown in  Fig. \ref{TBfig}.
 
 Note the important result from Eq. (\ref{obck}) that the orbital Berry curvature near the valley points are directly proportional to the magnitude of the respective intrinsic orbital moment 
 $M_z (\vec q)$. Furthermore, for the valence band tops at the two different valleys ($\tau = \pm 1$ and $\nu \tau = 1$; $\nu$ reverses sign as the valence band top is spin $\uparrow$ at $K$, while being spin $\downarrow$ at $K'$),
the orbital Berry curvature has exactly the same magnitude as well as the same sign at the 
 two valleys as dictated by the time reversal symmetry. 
Finally, Eq. (\ref{obck}) also shows that the two spin split valence bands ($\nu = \pm 1$) at each valley have slightly different orbital Berry curvatures, if the SOC parameter $\lambda$ is non-zero.
%


 The orbital Hall conductivity may be computed from the BZ sum of the orbital Berry curvature, using Eqs. (\ref{OHC}) and (\ref{obck}), over the occupied states, and results 
 will be given for both the doped and the undoped case.
 
The fact that the valley regions have the dominant contribution to the OHC can be argued from the 
energy denominator in the Kubo formula  (\ref{obc}) and this is also seen from the TB results presented in Fig. \ref{fig3} (b). In our results derived here, we therefore consider only the two valleys
$K$ and $K'$. 
  An analytical result is obtained by approximating circular Fermi surfaces  centered at the
 valley points. This is reasonable because the dominant contribution to the OHC
 comes from the valley points, quickly falling off as one goes away from the valley points, so that it's an excellent approximation to replace the actual Fermi surface 
 by circles around the valley points with the same total Fermi surface area.

 \begin{figure}[t]
\centering
\includegraphics[scale=.35]{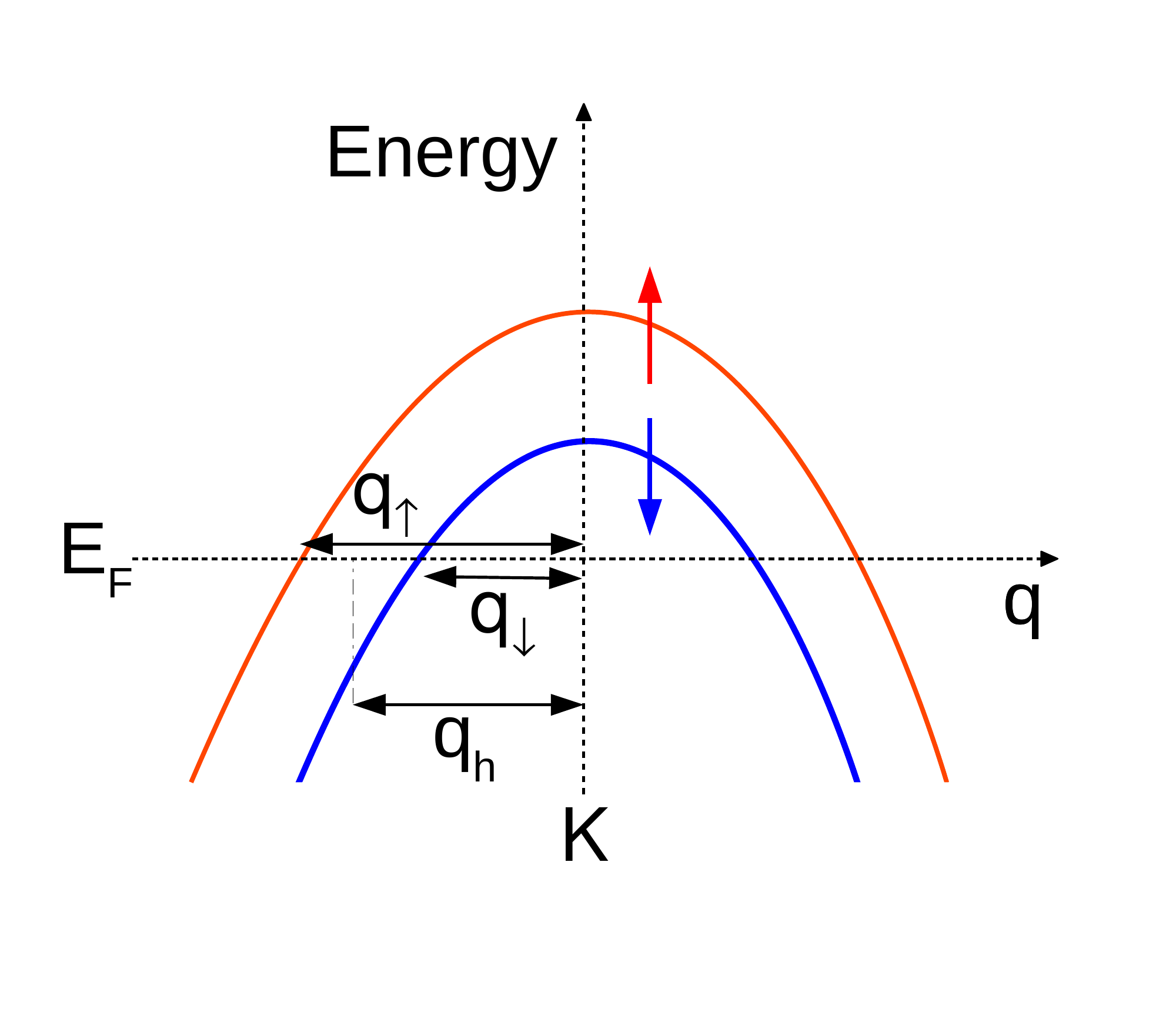}
\caption {The hole pocket and the spin-dependent hole pocket radii,
$q_\uparrow$ and  $q_\downarrow$, at the $K$ valley ($\tau = 1$) used in the 
integration to evaluate the  OHC and SHC expressions, Eqs. (\ref{sigma-orb}) and  (\ref{shc}). 
}
\label{figSHC}   
\end{figure}

The OHC can be computed by performing 
 the momentum sum analytically by integrating over the occupied states with the help of
 Fig. \ref{figSHC}. The hole pocket radii at the valley points are determined by the spin-orbit splitting
 and the dopant hole concentration $n_h$. Denoting the electron concentration by $n_e$, we have $n_h + n_e =2 $ for the valence bands. 
Considering that the OHC depends only weakly on the SOC parameter $\lambda$, the integration is simplified
by taking the limits from $q_h$ to $q_F$, where 
 $  \pi q_{\rm h}^2 = (n_h/4) A_{BZ}$, and $  \pi q_{\rm F}^2 = A_{BZ}/2$,
 where $A_{BZ}$ is the BZ area  and the factor of two in the last expression is due to the presence of two valley points in the BZ.
Putting these together, the expression (\ref{OHC}) for the OHC leads to the result 
 \begin{eqnarray}  \nonumber             \label{sigma-orb}
&&  \sigma^{z, orb}_{yx}  
 \approx    -\frac{2 e} { (2\pi)^2} \sum_{\nu = \pm 1} \int_{q_h}^{q_F} d^2 q  \times \Omega^{z,\rm orb}_{\nu,yx} ({\vec  q}) \\ \nonumber
&=& \frac{-e} {\pi} \times \Big[ \frac{\Delta}{(\Delta^2 + Aq_h^2)^{1/2}}
  -  \frac{\Delta}{(\Delta^2 + A q_F^2)^{1/2}} 
  \Big]
 + O( \frac {\lambda^2} {\Delta^2}),\\
 \end{eqnarray}
 where $A = 4t^2a^2$, and a factor of two is included to take into account the two valleys, which contribute
 the same amount. 

For the undoped case $q_h = 0$ , and plugging in the value  $q_F  = (A_{BZ}/(2 \pi))^{1/2}$,
 $A_{BZ} =  8 \pi^2/ \sqrt 3 a^2 $, we find the OHC to be \cite{Error}
  \begin{eqnarray}          \label{undoped} \nonumber 
  \sigma^{z, orb}_{yx} = \frac{-e} {\pi} \times \Big[ 1- \frac {\Delta}   
   {\sqrt { \Delta^2 +  16 \pi t^2/ \sqrt 3
  } } \Big] + O( \frac {\lambda^2} {\Delta^2}).\\
 \end{eqnarray}
%
 \begin{figure}[t]
\centering
\includegraphics[scale=.30]{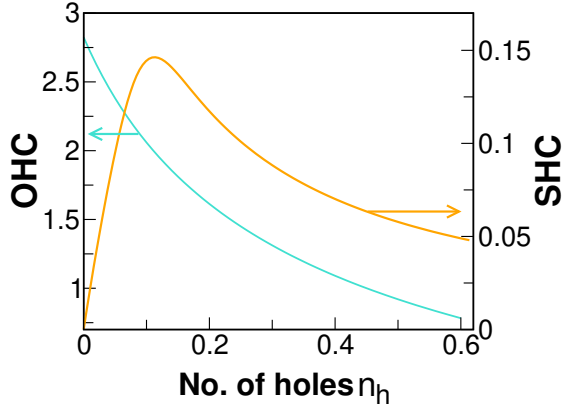}
\caption {Hall conductivities, OHC ($-\sigma^{z,\rm orb}_{yx}$) and SHC ($\sigma^{z,\rm spin}_{yx}$), both in units of $e/4\pi$, in the four-band model as a function of the hole concentration $n_h$, computed from 
Eqs. (\ref {sigma-orb}) and (\ref{shc}), respectively.
}
\label{4band-conductivity}   
\end{figure}
Clearly, in the undoped case the magnitude of the OHC is larger compared to the hole doped case in Eq. (\ref{sigma-orb}) and with increase in hole doping OHC decreases.
The computed OHC as a function of hole doping  is shown in Fig. \ref{4band-conductivity} together with the SHC. This qualitatively explains the tuning of OHC with hole doping, as obtained from the full TB calculations (Fig. \ref{fig_ohc}). However, in the four-band model, the magnitudes of both OHC and SHC are underestimated due to the overestimation
of the energy denominator in the Kubo formula (\ref{obc}) as one moves away from the valley points. 

\subsection {Effect of warping on the OHC} \label{warping} 
The OHC results calculated above within the four-band model do not include the trigonal warping of the band structure in real materials, although such warping effects are already present both in the tight-binding results of Section \ref{TB} as well as in the DFT results presented in Section \ref{SecDFT} later. In the interest of keeping the four-band model as simple as possible, it is desirable not to include the
warping of the band structure, as we have done above in the four-band model, which allows for the derivation of several important analytical results presented above. In this subsection, we examine the effect of the warping term in the band structure, and we find that for typical warping strength found in real TMDCs, the warping does not affect the basic physics of OHE, as discussed earlier.

\begin{figure}[t]
\centering
\includegraphics[scale=.25]{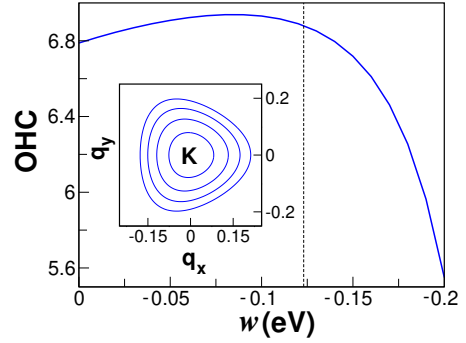}
\caption {
Dependence of the OHC ($-\sigma^{z,\rm orb}_{yx}$) on the warping of the band structure (warping parameter $w$), as obtained from the model Hamiltonian (\ref{TW}). 
Parameters typical for TMDCs were used: $t$, $\Delta$, and $\lambda$ are the same as in Fig. \ref{TBfig}, while the warping parameters were taken from Ref. \cite{3band}
, viz., 
$\gamma_1 = 0.077$ eV, $\gamma_2 = 0.055$ eV, and $w = -0.123$ eV.
The vertical dashed line corresponds to this typical value of $w $. The inset shows the constant energy contours for the topmost valence band with warping as obtained from Eq. (\ref{EqTW}).  
OHC is in units of $10^3 (\hbar/e) \Omega^{-1}$.
}
\label{fig_warping}   
\end{figure}

When the warping is present in the band structure, the energy bands in the vicinity of the valley points are no longer isotropic, but they acquire a directional dependence, distorting the circular contours, 
as indicated in the inset of Fig. \ref{fig_warping}. 
The effect may be described by adding a warping term ${\cal H}_w (\vec q )$, well known in the  literature \cite{Kormanyos,3band}, to the   
valley 
 Hamiltonian ${\cal H}_v (\vec q )$ of Eq. (\ref{HK}).
This additional term in our standard basis ($ |u \rangle, |d \rangle, $) reads  
 \begin{eqnarray}    \nonumber 
{\cal H}_w (\vec q ) &=&  
\left [
{\begin{array}{*{2}c}
    \gamma_1 q^2a^2  &  w(\tau q_x -iq_y)^2a^2\\  
     w(\tau q_x + iq_y)^2a^2 & \gamma_2 q^2a^2  
   \end{array} }  \right]. \\
 \end{eqnarray} 
Here, the new parameters $\gamma_1$ and $\gamma_2$ take into account  the asymmetry of the conduction and valence bands in the band structure, while $w$ is the warping parameter, representing the   anisotropic dispersion, known as ``trigonal warping'' of the energy bands, in the neighborhood of the valley points. These additional parameters may be expressed, if desired, in terms of the 
tight-binding hopping integrals similar to Eq. (\ref{TB-parameters}). The full Hamiltonian with the warping and SOC terms included then becomes
\begin{equation}    \label{TW}
{\cal H}^{\rm SOC}_w (\vec q ) =  {\cal H}^{\rm SOC}_v (\vec q ) + {\cal H}_w (\vec q ) \otimes I_s,
\end{equation}
where again $I_s$ is the $2 \times 2$ identity matrix in the spin space, 
and ${\cal H}^{\rm SOC}_v (\vec q ) $ is the valley point Hamiltonian, Eq. (\ref{HKSOC}),  without the warping term.

%

A straighforward diagonalization yields the four eigenvalues
\begin{eqnarray}     \nonumber
 && \varepsilon^\nu_{\pm} = 2^{-1} \Big[ (\gamma_+ q^2a^2 
+ \nu\tau\lambda  \ ) \pm  
  \Big \{  8\tau a^3 t w  (q_x^2-3q_y^2)  q_x  + \\              
 &&   (\gamma_-^2 + 4w^2)q^4a^4  + \Delta_\nu^2 
+ 4t^2q^2a^2 -2 \gamma_-  \Delta_\nu q^2a^2  
 \Big \}^\frac{1}{2} \Big],
 \label{EqTW}
\end{eqnarray}
where the direction dependence of the band structure in the momentum space is clearly seen.
Here, $\gamma_\pm \equiv \gamma_1 \pm \gamma_2$,
$\Delta_\nu \equiv  \Delta -\nu \tau \lambda $, and in $\varepsilon^\nu_{\pm}$, the subscript $\pm$ sign denotes the valence and the conduction bands, while $\nu = \pm 1$ denotes the spin-split bands within the conduction or the valence band manifolds.
The constant energy contours for the topmost valence band around the $K$ valley with the warping included are shown in the inset of   Fig. \ref{fig_warping}. 

The OHC in the presence of warping is obtained from the eigenvalues and the eigenfunctions and using the Kubo expression, Eqs. (\ref{OHC}) and (\ref{obc}). We have not attempted to obtain the results analytically with the warping term present, but rather satisfied ourselves by numerically computing the results for typical parameters. 
These results are shown in Fig. \ref{fig_warping} for the undoped, insulating case and typical parameters  have been used \cite{3band}.
As seen from the figure (where $\gamma_1$ and $\gamma_2$ are fixed, but $w$ is varied), we find that for the typical parameter $w = 0.12$ eV, 
OHC does not change very much from the unwarped case ($w = 0$). 

For the Hamiltonian parameters $t = 1.02 $, $\Delta = 1.69$, $\lambda = 0.02$, $\gamma_1 = 0,077$, $\gamma_2 = 0.055$, and $w = -0.123$ (all in units of eV),
we compare the magnitude of the OHC both with and without
 ($\gamma_1 = \gamma_2 = w = 0$)
 the warping term
in the Hamiltonian (\ref{TW}). The results are $\sigma^{z, orb}_{yx} = -5.5 $ without
and -6.9 with warping, in units of $10^3 (\hbar / e) \ \Omega^{-1}$,
both of which are 
smaller than the DFT results ($\sigma^{z, \rm orb}_{yx} \approx -10$)
 presented in the next Section.
This is, as already stated,  due to the overestimation of
the energy denominator in the Kubo formula (\ref{obc}).


\subsection {Spin Hall Effect } 

We now turn to the calculation of the SHC within the four-band model described by
the Hamiltonian (\ref{HKSOC}).
 We have not included the warping term described in the previous subsection.
 Following the same procedure discussed in Section \ref{soc},
 we first compute the spin Berry curvature, which yields the result
\begin{eqnarray}            \label{sbck}
 \Omega^{z,\rm spin}_{\nu, yx} (\vec q) 
=  \frac{ \nu M_z (\vec q)}
 { \Delta+    \lambda (\nu \tau -2) }
 = \frac{\nu\tau}{2}   \Omega^{z,\rm orb}_{yx} (\vec q).
\end{eqnarray}  
The second equality is obtained by comparing with the expression (\ref{obck}) for the 
orbital Berry curvature.

Note the important result that the orbital Berry curvature gives rise to the spin Berry curvature, simply because $\Omega^{z,\rm spin}_{\nu, yx} (\vec q)$ becomes zero in absence of $\Omega^{z,\rm orb}_{\nu, yx} (\vec q)$. Furthermore, for $\lambda = 0$, the two spin-polarized bands ($\nu \pm 1$) have exactly equal and opposite contributions, as seen from the middle expression in Eq. (\ref{sbck}),  leading to a net zero SHC. This is in contrast to the OHC, which is nonzero even when $\lambda = 0$ [see Eq. (\ref{sigma-orb})], indicating the fundamental nature of the OHC.

When the SOC term is present ($\lambda \ne 0$), Eq. (\ref{sbck}) shows that the two spin split bands ($\nu = \pm 1$) contribute with opposite signs, but the magnitudes don't cancel exactly, leading to negligibly small SHC as compared to the OHC. 
When both bands are fully occupied ($n_e = 2$), 
a straightforward integration over the Brilloun zone yields the result
\begin{equation}
 \sigma_{yx}^{z, spin} \sim  -e \lambda  (\pi \Delta)^{-1}.
\end{equation}

Note that this is only the valley point contribution and the total SHC should be  computed by considering the contributions from the other points of the BZ as well. This is crucial for the undoped case, as in this case, the net SHC is not dominated by the valley point contribution, but nevertheless the  magnitude of the SHC remains much smaller than the OHC.

The situation for the SHC is somewhat different for the hole doped case,
as in that case, both spin-split bands are not occupied at every k point (see Fig. \ref {figSHC}).
The SHC is dominated by the region in the momentum space where one of
the two spin split bands are occupied. For small hole doing, the holes go to the 
valley points, and only for much larger hole concentration $n_h$, the holes go to the $\Gamma$ point.

Assuming the hole occupancy only at the valley pockets,  the SHC
is computed by integrating the lowest valence band ($\nu \tau = -1$) over an annular ring with inner radius $q_\downarrow$ 
and outer radius $q_\uparrow$, as shown in  Fig. \ref{figSHC}.
The result is
\begin{eqnarray}            \label{shc}
 \sigma^{z,\rm spin}_{\nu, yx} 
\approx \frac{e}{4\pi} \Big[ \frac{1}{(1+Bq_\downarrow^2)^{1/2}} - \frac{1}{(1+Bq_\uparrow^2)^{1/2}}\Big],
\end{eqnarray} 
where $ B =4t^2a^2/(\Delta + \lambda)^2$. 
In the very small hole doping limit, $q_\downarrow = 0$ and $q_\uparrow  = q_h$. In this limit, it is clear from Eq. (\ref{shc}) that the SHC increases with $n_h$. After a critical hole concentration, the Fermi surface forms the annular region, as depicted in Fig. \ref{figSHC}, and the SHC starts to decrease with $n_h$ (see Fig. \ref{4band-conductivity}). This explains the qualitative behavior of SHC in presence of hole doping as obtained from our numerical results in  Section \ref{soc} (see Fig. \ref{fig_ohc}). 

An important point to note from  Fig. \ref{4band-conductivity} is that with hole doping, while the SHC increases significantly from its near zero value, its magnitude nevertheless remains much smaller than the OHC for all $n_h$, making the TMDCs to be prime candidates for the experimental observation of the OHE.

 \begin{figure*}[t]
\centering
\includegraphics[scale=.32]{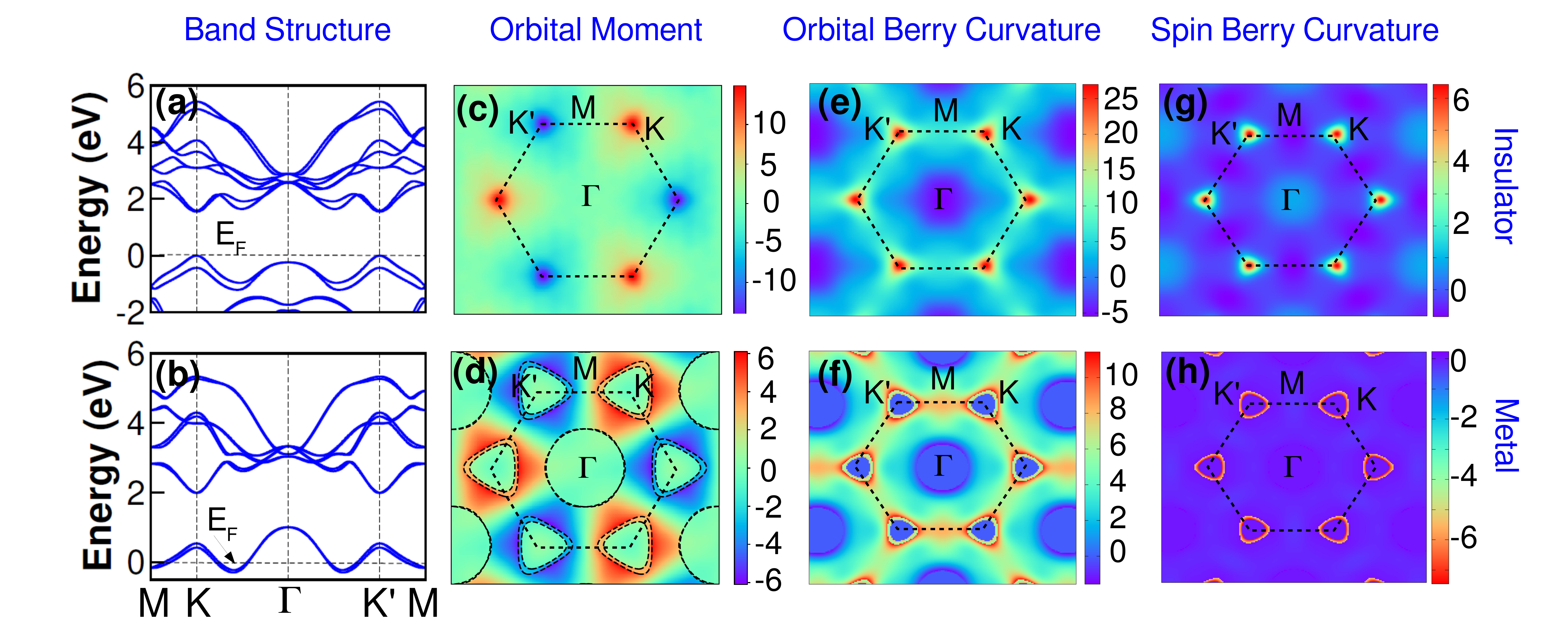}
\caption {DFT results for the prototypical insulating and metallic monolayer TMDCs with the SOC included. (a) Band structures for insulating WS$_2$. (b) Band structure for the metallic NbS$_2$. 
(c) Orbital moment $M_z(\vec k)$ (eV$\cdot$\AA$^2$) for WS$_2$, 
summed over all occupied bands (including the lower lying $p$ bands), in the $k_x$-$k_y$ plane. 
(d) The same  for NbS$_2$. The hexagons on all figures indicate the Brillouin zone. The circular dashed lines in (d) indicate the Fermi surface, with the hole pockets centered
around the $\Gamma$, $K$, and the $K'$ points. The $p$ bands contribute to $M_z(\vec k)$ very little, and therefore $M_z(\vec k)$ is nearly zero in the hole pockets.
(e) and (f) Orbital Berry curvatures summed over the occupied bands at each $k$ point (in units of \AA$^2$) for  WS$_2$ and NbS$_2$, respectively. (g) and (h) The  corresponding spin Berry curvature sums for the two materials  (in \AA$^2$). For the NbS$_2$ case (h), the dominant contribution to SHC comes from the
red annular regions  around $K$, $K'$ points, where just one spin band is occupied. 
Everywhere else in (h), it is nearly zero. As discussed in the text, occupation of both spin bands at any $k$ point results in a near cancellation of the contribution to the SHC (which is the spin Berry curvature sum) originating from that point.
}
\label{fig6} 
\end{figure*}

\vspace{1 cm}

\section{Density functional results} \label{SecDFT}

While the model calculations, discussed above, provide an understanding of the physics of the OHE, in order to compute the magnitude of this effect for real materials, we have performed density functional  calculations of the OHE and the SHE for a series of TMDCs.
Results are presented for both insulating materials, MX$_2$ (M = Mo, W; X = S, Se, Te), as well as 
for the metallic  NbS$_2$. 
The results show, as anticipated from the model calculations, that while both insulating and metallic materials have a strong OHE, the SHE is relatively weak in all materials, and negligibly small in the insulating TMDCs.

\subsection{Insulating  TMDCs, MX$_2$ (M = Mo, W;  X = S, Se, Te)} 

In Mo and W based monolayer TMDCs MX$_2$, the metal atom is in the $d^2$ configuration, 
and these materials exhibit an insulating band structure, with a  direct band gap at the valley points. 
The band structure of WS$_2$ is shown in Fig. \ref{fig6} (a).
As already discussed, two key features of the band structure are the valley-dependent complex orbital characters 
and the spin splitting of the valence bands due to SOC.  
The former leads to the orbital moments at the two valley points with opposite signs that eventually results in a large OHC, while the latter leads to a non-zero SHC, though it is much smaller in magnitude.

 The computed magnitudes of the OHC and SHC for a series of insulating TMDCs as well as a 
 representative metallic TMDC are listed in Table \ref{tab2}, and the key results of the DFT calculations
 are shown in Fig. \ref{fig6}.
 As seen from Table \ref{tab2} and
Fig. \ref{fig6} (e), the dominant contributions to OHC come from the region near the valley points. 
Also, the contributions from the region near the $\Gamma$ point appear with an opposite sign and have a much smaller magnitude. Given the fact that the orbital moment at the $\Gamma$ point is zero (see Fig. \ref{fig6} (c)), this small contribution to OHC, as listed in Table \ref{tab2}, can be attributed to the electric field induced orbital texture, similar to the centrosymmetric materials reported earlier \cite{Go, Jo}.

As is clear from the discussions on the four-band model, Section \ref{b},  
OHE gives rise to SHE, when the SOC is present.
As seen from Table \ref{tab2}, the calculated SHC has a much smaller magnitude than the OHC, viz., $\sigma^{\rm orb}/\sigma^{\rm spin} \approx 10^3$, consistent with our NN model calculations.
The reason for the small SHC, again, is that if both spin up and down bands are occupied at a particular $k$ point in the BZ, they nearly cancel the contributions to the SHC at that $k$ point. 
As might be expected, the DFT results show that the larger the magnitude of the SOC parameter $\lambda$, the larger is the SHC, even though in all cases they remain much smaller than the OHC.


We note that in centrosymmetric 3$d$ transition metals, such as V, Cr, Mn, considered earlier \cite{Jo},
the SHC was small simply because the SOC parameters are small in $3d$ metals, unlike the insulating TMDCs, where SHC is small due to the occupation of both spin bands at each $k$ point. This
led to a large ratio
 $\sigma^{\rm orb}/\sigma^{\rm spin} \approx 65-70$ for the 3$d$ transition metals\cite{Jo}, which is much smaller as compared to the TMDCs, where this ratio is $\approx 10^3$. This suggests the insulating TMDCs to be good materials for 
 experimental observation of the OHE.

\begin{table*} [t!]    
\caption{DFT results for the OHC and the SHC of insulating and metallic monolayer TMDCs. The partial contributions to OHC ($\sigma^{z,orb}_{yx} = \sigma_{\rm K} + \sigma_{\rm \Gamma} + \sigma_{\rm rest}$) are also listed, where
$\sigma_{\rm K}$, $\sigma_{\rm \Gamma}$, and $\sigma_{\rm rest}$ are the contributions, respectively, from the valley, $\Gamma$-point, and the remaining regions of the BZ. 
The SHC are all small, but increase with the SOC parameter $\lambda$. 
It is larger for the metallic  NbS$_2$ because of the non-cancellation of the contributions 
from the two spin bands at the edge of the hole pockets as discussed in the text.
%
}

\centering
\setlength{\tabcolsep}{7pt}
 \begin{tabular}{c|c | c c c c  |c  c}
\hline
 & Materials & \multicolumn{4}{c|}{OHC [in $10^3 \times (\hbar/e)   
 \Omega^{-1}$]} & \multicolumn{2} {c}{SHC in $(\hbar/e)   
 \Omega^{-1}$}\\
 & & $\sigma_{\rm K}$ & $\sigma_{\rm \Gamma}$ & $\sigma_{\rm rest}$ & $\sigma^{z,orb}_{yx}$ & $\lambda$ (eV) & $ \sigma^{z, spin}_{yx}$\\
\hline
\multirow{6}{*}{\begin{turn}{90}Insulator \end{turn}} & WS$_2$ & -7.7 & 1.4 & -3.7 & -10.0 & 0.21 & 5.2\\
& WSe$_2$ & -8.5 & 1.4 & -3.0 & -10.1 & 0.23 & 7.1\\
 & WTe$_2$ & -8.6 & 1.0 & -2.6 & -10.2 & 0.24 & 9.4 \\
& MoS$_2$ & -9.1 & 1.7 & -3.2 & -10.6 & 0.08 &1.0\\
 & MoSe$_2$ & -8.0 & 1.7 & -3 & -9.3 & 0.09  &1.8\\
 & MoTe$_2$ &-9.1  & 1.1 & -2.5 & -10.5 & 0.11 & 3.0\\
 \hline 
\multirow{2}{*}{\begin{turn}{90} Metal \end{turn}} & \multirow{2}{*}{NbS$_2$} & \multirow{2}{*}{-5.8}  & \multirow{2}{*}{0.04} & \multirow{2}{*}{-3.7} & \multirow{2}{*}{-9.5} & \multirow{2}{*}{0.06} & \multirow{2}{*}{367}\\
 & &   &  &  &  & \\
\hline
\end{tabular} 
\label{tab2}
\end{table*}

\subsection{Metallic TMDC (NbS$_2$)}

We have considered the monolayer NbS$_2$, as a metallic counterpart
of the TMDCs.
Here Nb is in the $d^1$ electronic configuration, as a result of which the system becomes metallic, with the corresponding band structure shown in Fig. \ref{fig6} (b). 

The calculated orbital moment as well as the  Fermi surface are shown in Fig. \ref{fig6} (d), where we can see a circular hole pocket around the $\Gamma$ point and a 
triangular hole pocket around each valley point. 
Each hole pocket has actually a slightly different boundary for the two spins 
(most clearly visible around the valley points) due to the spin orbit split bands. 
As seen from the figure, due to the unoccupied $d$ states in the vicinity of the valley points (hole pocket), the orbital moment decreases substantially. 
This, in turn, results in a smaller magnitude for the orbital Berry curvature, as depicted in Fig. \ref{fig6} (f). As a result, the  OHC in NbS$_2$ is much smaller as compared to the insulating TMDCs (see Table \ref{tab2}).

In contrast to the insulating TMDCs, NbS$_2$ has a large spin Berry curvature [see Fig. \ref{fig6} (h)] and hence a larger SHC. This is in contradiction to the common belief that SHC is larger in materials with larger SOC, as the strength of SOC in NbS$_2$ is smaller than the  other Mo and W based TMDCs (see Table \ref{tab2}). 

The larger SHC in NbS$_2$  can be understood by comparing its band structure with that of the insulating TMDCs. Unlike the insulating TMDCs, in NbS$_2$, the two spin-polarized valence bands near the valley points are only partially occupied. 
As a result, the region in the momentum space, where only one of the two spin split valence bands 
is occupied [see Fig. \ref{fig6} (h)], gives a large contribution to the SHC.
In the insulating TMDCs, as discussed earlier, the 
occupation of the two spin-split bands leads to a near cancellation of the
spin Berry curvature at every k point, leading to a very small SHC.
The physics is the same as that for the hole doped insulators 
illustrated in Section \ref{b} and Fig. \ref{4band-conductivity}.

\section{Summary}

To summarize, we studied the orbital Hall effect in a family of TMDC materials, with broken inversion symmetry, from both density-functional calculations and model studies. Both insulating and metallic systems were considered and the effects of hole doping as well as band structure changes were also studied. 
Explicitly, we studied the insulating materials MX$_2$ (M = Mo, W; X = S, Se, Te) and the 
prototype metallic compound NbS$_2$. 
The spin Hall effect was also calculated, and the magnitude of the spin Hall conductivity was found to be extremely small as compared to the orbital Hall conductivity in all cases, suggesting this class of materials 
to be excellent candidates for the experimental study of the orbital Hall effect.

The physics of the OHE in the TMDCs is governed by the presence of a robust intrinsic orbital moment in the 
momentum points in the Brillouin zone, which is non-zero due to the broken inversion symmetry. 
The effect is furthermore dominated by the two valley points $K$ and $K'$ in the Brillouin zone,
for which we developed a minimal, four-band model, which allowed for an analytical solution providing
considerable insight into the basic physics of the OHE and the SHE in these materials.

While the spin-orbit coupling is not necessary for the OHE, which exists even without it,
the presence of the SOC term is essential for a non-zero SHE. In this sense, the OHE is
more fundamental than the SHE. While the OHE has a very weak dependence on the SOC,
the SHE depends very strongly on the SOC term, but still always being considerably weaker in magnitude 
than the OHE, even for metallic or hole doped systems.
In this context, even though the relatively large SHE in NbS$_2$ as compared to the other TMDCs studied
looks contradictory at a first glance  due to the small value of $\lambda$ in NbS$_2$, such behavior can be
understood from 
 the spin-split band structure, where occupation of only one of the two spin-split bands
 in a small part of the Brillouin zone near $E_F$ leads to a strong contribution to the SHE, even though
 $\lambda$ is small.

In spite of the fact that the OHE has been suggested earlier from theoretical considerations in order to explain the large spin Hall and anomalous Hall effects in certain materials\cite{Kotani2009,Tanaka,Kotani2008,Jo}, the effect has not yet been 
established from direct experimental measurements.  
Our work suggests the TMDCs to be good systems to observe the OHE.
The OHE can be probed in experiments like magneto-optical Kerr measurements that can be used to detect the orbital moments accumulated at the edges of the sample due to the OHE \cite{Kato}. The OHE can also be measured by detecting the orbital torque generated by the orbital Hall current \cite{Go2019}. Furthermore, 
photon polarized angle-resolved photoemission measurements  \cite{Park} can be used to detect the proposed valley-orbital locking, which is analogous to the well known valley-spin locking.

The intrinsic orbital moments leading to the   strong OHE is in essence a consequence
of the broken inversion symmetry,  and as such the ideas presented here should not only be applicable to the TMDCs, but rather to all 2D materials with broken symmetry. 
Once established, the OHE in 2D materials can open up new avenues for future research and potential applications in orbitronics devices.

\begin{acknowledgements}

 We thank the U.S. Department of Energy, Office of Basic Energy Sciences, Division of Materials Sciences and Engineering for financial support under Grant No. DEFG02-00ER45818.
\end{acknowledgements} 

{
\vspace {2 cm}
\centering
{\bf {\center APPENDIX A: L\"owdin downfolding and Inter-orbital hopping due to broken inversion symmetry}}
}
\vspace {.2 cm}

In this Appendix, we illustrate how the broken inversion symmetry can introduce additional hopping terms between neighboring atoms by taking an example relevant to the TMDCs.

\begin{figure}[h] 
\centering
\includegraphics[scale=0.25]{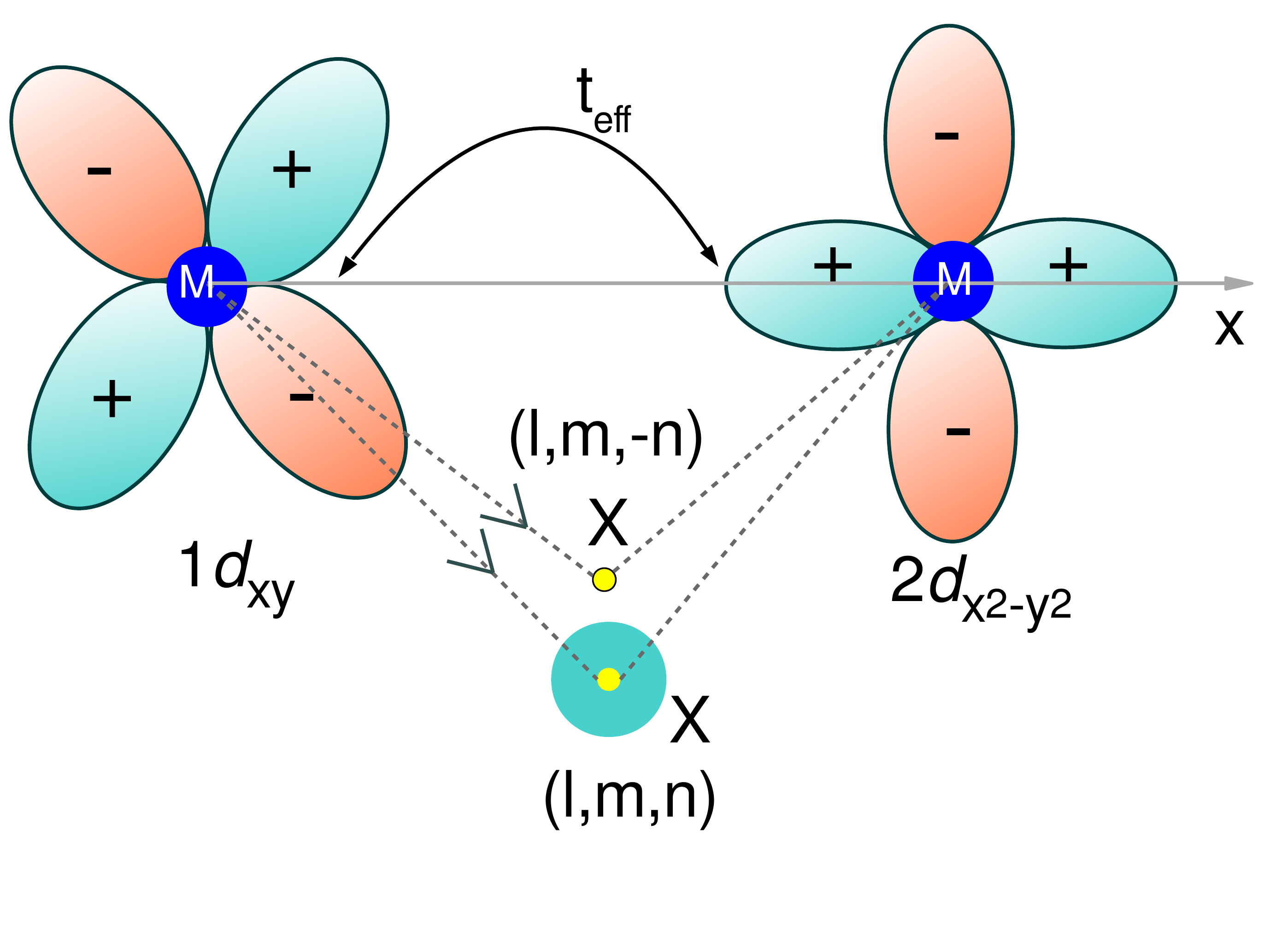}
\caption {Illustration of the modification of the inter-orbital hopping between the two metal atoms caused by the broken inversion symmetry due to the presence of the two ligand atoms $X$ with the direction cosines $(l, m, \pm n)$, measured with respect to the origin at the left metal atom. Here inversion symmetry is broken, but not the mirror symmetry $\sigma_h$.
}
\label{fig-Appendix}  
\end{figure}

Consider the hopping  between the two $d_{xy}$ and $d_{x^2-y^2}$ orbitals along $\hat x$ as indicated in Fig. \ref{fig-Appendix}. From symmetry, this hopping is zero without the presence of the ligand atoms $X$.
When the ligand atoms are introduced, the orbital lobes of the metal atoms change in a non symmetric manner,
and two things happen: (a) The hopping  is no longer zero and (b) It changes sign for hopping in the opposite direction. 
Mathematically, this is described via the L\"owdin downfolding\cite{downfolding},
where the ligand atoms are completely removed and their effect is folded into effective hopping integrals 
($t_{\rm eff}$) between the
metal orbitals. This is illustrated in this Appendix.

For simplicity, we consider the symmetry broken by the ligand X $s$ orbitals, situated at the bottom ($s_b$) and top ($s_t$) of the plane. The same arguments can be easily generalized   for the X $p$  orbitals as well.


Using the Slater-Koster Tables \cite{SlaterKoster}, we can write the hopping matrix in the basis set $\phi_\alpha \equiv \{ |1 d_{xy} \rangle,|2 d_{x^2-y^2}\rangle, |s_t\rangle, |s_b\rangle \}$,  

\begin{widetext}

\begin{eqnarray}     \label{Hhop}  
H =
\left[
{\begin{array}{*{20}c}
    \epsilon_d & 0 & \sqrt{3}lm V_{sd\sigma} & \sqrt{3}lm V_{sd\sigma}   \\
    0 & \epsilon_d & \sqrt{3}(l^2-m^2) V_{sd\sigma}/2 & \sqrt{3}(l^2-m^2) V_{sd\sigma}/2\\
    \sqrt{3}lm V_{sd\sigma} &  \sqrt{3}(l^2-m^2) V_{sd\sigma}/2 & \epsilon_s & V_{ss\sigma}\\
    \sqrt{3}lm V_{sd\sigma} & \sqrt{3}(l^2-m^2) V_{sd\sigma}/2   & V_{ss\sigma} & \epsilon_s\\
   \end{array} }  \right]
  \equiv 
   \left[
   {\begin{array}{*{20}c}
    h &  b  \\
    b^\dagger & c\\
       \end{array} }\right].
    \end{eqnarray} 

\end{widetext}
 Here ($l, m, \pm n$) denote the direction cosines of the top and bottom $s$ orbitals with respect to the M-$d_{xy}$ orbital at site 1 (left metal atom). Note that the direction cosines of the $d_{x^2-y^2}$ orbital at site 2 (right metal atom) is (1 0 0) with respect to the M-$d_{xy}$ orbital at site 1, which results in a vanishing direct hopping between $d_{xy}$ and $d_{x^2-y^2}$ orbitals, and  $\epsilon_s$ and $\epsilon_d$ are the onsite energies.
 
 In Eq. (\ref{Hhop}), we have separated the hopping matrix $H$ into four blocks. The block $h$ contains the important $d$ orbital space, and if the block $c$ is well separated in energy ($\epsilon_d  >> \epsilon_s$),  the effect of the unimportant
 $c$ block can be folded into the $d$ orbital space, which would produce an effective hopping matrix $h_{\rm eff}$.
 The result for $h_{\rm eff}$ is given by the well-known L\"owdin downfolding equations \cite{downfolding}.
The expression is
\begin{eqnarray} \nonumber\label{DF}
 h^{\rm eff} &=& h + b(\varepsilon I -c)^{-1} b^\dagger, \\
 h^{\rm eff}_{ij} &=& h_{ij} +\sum_k \frac{b_{ik} (b^\dagger)_{kj}}{h_{ii}-c_{kk}},
\end{eqnarray}
where the last step represents the lowest order in Brillouin-Wigner perturbation series \cite{BW}, obtained from the iterative solution and is valid since $b_{ik} << |h_{ii} - c_{kk}|$.
After some straightforward algebra, Eq. (\ref{DF}) yields the desired result
\begin{eqnarray}     \label{hop}  
h^{\rm eff}  &=&                                                 
\left[
{\begin{array}{*{80}c}
\epsilon_d + 6l^2 m^2 \epsilon_{sd} & t_{\rm eff}  \\
t_{\rm eff}  & \epsilon_d + \frac{3}{2} (l^2-m^2)^2 \epsilon_{sd} \\
   \end{array} }  \right],
    \end{eqnarray} 
where 
\begin{eqnarray} 
t_{\rm eff} = 3lm (l^2-m^2)\epsilon_{sd},
\label{tdd}
\end{eqnarray}
with $\epsilon_{sd}  =  V^2_{sd\sigma} / (\epsilon_d  - \epsilon_s ) $.
Thus, a non-zero hopping $t_{\rm eff}$ between the $d_{xy}$ and the $d_{x^2-y^2}$ orbitals is introduced
due to the broken inversion symmetry, which is the first item we set out to show in this Appendix. 

The second result is the change of sign of $t_{\rm eff}$ with the direction of hopping. 
Following the same logic as above, it can be shown that if the second metal atom is shifted from $\hat x$ to 
$- \hat x$, along with the ligand atoms, Eqs. \ref{hop} and \ref{tdd}   still hold except that the direction cosine $l$ is changed to $-l$, so that the effective hopping $t_{\rm eff}$
changes sign.

{
\vspace {1 cm}
\centering
{\bf {\center APPENDIX B: Broken Inversion Symmetry and Orbital Moment}}
}
\vspace {.2 cm}

The sign change in the effective hopping $t_{\rm eff}$ plays a crucial role in generating the intrinsic orbital moment. In this Appendix, we illustrate how  the sign change leads to the formation of complex orbitals, which, in turn, carry a nonzero orbital moment. 

To illustrate this point, we consider a simple one-dimensional lattice model with two orbitals ($\alpha = d_{xy}$ and $d_{x^2-y^2}$)
on each atom. It is easy to see that the inversion symmetry breaking  does not alter the sign of the NN hopping between two similar orbitals, while the sign is changed between two dissimilar orbitals when the hopping direction is reversed. Denoting the NN hoppings as:
$\langle \alpha | H | \alpha \rangle_{\pm \hat x} =   t_0 $, 
$\langle d_{x^2-y^2} | H | d_{xy}  \rangle_{ \hat x}=
\langle d_{xy} | H | d_{x^2-y^2} \rangle_{-\hat x}
= t_{\rm eff},$
and $\langle d_{x^2-y^2} | H | d_{xy}  \rangle_{-\hat x}=
 \langle d_{xy} | H | d_{x^2-y^2} \rangle_{\hat x}  =
 - t_{\rm eff} $,
where the subscript denotes the displacement of the second orbital with respect to the first.

The TB Hamiltonian in the momentum space 
is
\begin{eqnarray}     \label{hopk}  
H (k )  &=&                                                 
\left[
{\begin{array}{*{80}c}
\epsilon_d^\prime (k) & 2it_{\rm eff} \sin k a \\
-2it_{\rm eff} \sin k a  & \epsilon_d^\prime (k)   \\
   \end{array} }  \right]
    \end{eqnarray} 
which is written in the Bloch function basis 
$a^\dagger_{k \alpha} = N^{-1/2} \sum_{n=1}^N e^{ikna} c^\dagger_{n \alpha}$, where $c^\dagger_{n \alpha}$
creates an electron at site $n$ in the orbital $\alpha$, and $\epsilon_d^\prime = \epsilon_d +  2t_0 \cos(k a)$.
The energy eigenvalues are $E_{\pm} = \epsilon_d^\prime  \pm 2t_{\rm eff} \sin k a$, 
but more interestingly, the corresponding wave functions mix the two orbitals at each site with complex coefficients, viz., 
$|d_{x^2-y^2}\rangle \pm i|d_{xy}\rangle $, which carries the 
orbital moment $L_z = \pm 2 \hbar$. 
This illustrates the role of the  broken $\mathcal{I}$ symmetry in generating the intrinsic orbital moment
at various points in the Brillouin zone.
%

\begin{widetext}

{
\centering
{\bf {\center APPENDIX C: Character Table of the $D_{3h}$ point group}}
}
\vspace {.2 cm}

The crystal structure for the MX$_2$ TMDC in the 2H structure has the $D_{3h}$ symmetry,
which lacks the inversion symmetry, but contains the mirror symmetry $\sigma_h$, which has important 
implication for the electronic structure as discussed in the text.

 \begin{table*}  [h]   
\caption{Character Table of $D_{3h}$ and the irreducible representations spanned by the metal $d$ orbitals.}
\centering
\setlength{\tabcolsep}{12pt}
 \begin{tabular}{ c| c c c c c c |c}
\hline\hline
$D_{3h}$ \ & E \ & 2$C_3$ & 3$C_2'$ & $\sigma_h$ & 2$S_3$ & 3$\sigma_v$ & \\
\hline
$A_1'$ & 1 & 1 & 1 & 1 & 1 & 1 & $x^2+y^2$, $z^2$ \\
$A_2'$ & 1 & 1 & -1 & 1 & 1 & -1 & \\
$E_1'$ & 2 & -1 & 0 & 2 & -1 & 0 & $x^2-y^2$, $xy$\\
$A_1''$ & 1 & 1 & 1 & -1 & -1 & -1 & \\
$A_2''$ & 1 & 1 & -1 & -1 & -1 & 1 & \\
$E_1''$ & 2 & -1 & 0 & -2 & 1 & 0 & $xz$, $yz$\\
\hline\hline
\end{tabular} 
\label{characterTable} 
\end{table*}
\end{widetext}



\begin{thebibliography}{10}
\bibitem{Error} Equation (\ref{undoped}) corrects Eq. (11) of our earlier paper \cite{ohe_our}, where we had missed a factor of $\sqrt 2$ in the upper limit of the integration. All other equations and all numerical results and conclusions presented in our earlier paper remain unchanged.
\end{thebibliography}
\end{document}